\documentclass[journal]{IEEEtran}
\usepackage{cite}
\usepackage[pdftex]{graphicx}
\usepackage[cmex10]{mathtools}
\usepackage{amssymb}
\usepackage{epsfig}
\usepackage{subfigure}
\usepackage{algorithm}
\usepackage{algpseudocode}
\usepackage[usenames,dvipsnames]{color}
\usepackage{array}

\usepackage{bm}
\usepackage{bbm}
\usepackage{microtype}
\usepackage{kotex}
\usepackage{tabto}
\usepackage{makecell}
\usepackage{setspace}

\newtheorem{prop}{Proposition}
\newtheorem{remark}{Remark}

\def\ba{{\mathbf{a}}}  \def\bc{{\mathbf{c}}} \def\bd{{\mathbf{d}}}
\def\bee{{\mathbf{e}}} \def\bff{{\mathbf{f}}}  \def\bh{{\mathbf{h}}}
   
 \def\bn{{\mathbf{n}}}  
\def\bq{{\mathbf{q}}}  \def\bs{{\mathbf{s}}} 
\def\bu{{\mathbf{u}}} \def\bv{{\mathbf{v}}} \def\bw{{\mathbf{w}}} \def\bx{{\mathbf{x}}}
\def\by{{\mathbf{y}}} \def\bz{{\mathbf{z}}}  

\def\bA{{\mathbf{A}}}  \def\bC{{\mathbf{C}}} 
 \def\bF{{\mathbf{F}}}  \def\bH{{\mathbf{H}}}
\def\bI{{\mathbf{I}}}   
   
 \def\bR{{\mathbf{R}}}  
   
 \def\bZ{{\mathbf{Z}}}

\DeclareMathOperator*{\argmin}{arg\,min}
\DeclareMathOperator*{\argmax}{arg\,max}

\begin{document}


\onecolumn
\textbf{IEEE Copyright Notice}
\\
\\
© 2019 IEEE. Personal use of this material is permitted. Permission from IEEE must be obtained for all other uses, in any current or future media, including reprinting/republishing this material for advertising or
promotional purposes, creating new collective works, for resale or redistribution to servers or lists, or reuse of any copyrighted component of this work in other works.

\newpage

\twocolumn
\title{{Limited Feedback Designs for Machine-type Communications Exploiting User Cooperation}}


\author{Jiho Song,~\IEEEmembership{Member,~IEEE,} Byungju Lee,~\IEEEmembership{Member,~IEEE,} Song Noh,~\IEEEmembership{Member,~IEEE,} and Jong-Ho Lee,~\IEEEmembership{Member,~IEEE,}  
\thanks{J.\ Song is with the School of Electrical Engineering, University of Ulsan, Ulsan, South Korea (e-mail: jihosong@ulsan.ac.kr).}
\thanks{B.\ Lee \textit{(corresponding author)} is with Samsung Research, Seoul, Korea (email: byungjulee1730@gmail.com).}
\thanks{S.\ Noh is with the Department of Information and Telecommunication Engineering, Incheon University, Incheon, South Korea (email: songnoh@inu.ac.kr).}
\thanks{J.-H. Lee is with the School of Electronic Engineering, Soongsil University, Seoul, South Korea (e-mail: jongho.lee@ssu.ac.kr).}
\thanks{This work was supported by the 2018 Research Fund of University of Ulsan.}
}

\maketitle


\begin{abstract}
Multiuser multiple-input multiple-output (MIMO) systems are a prime candidate {for use in} massive connection density {in machine-type communication (MTC) {networks}.}
{One of the key challenges {of MTC networks} is to obtain accurate channel state information (CSI) at the access point (AP) {so that} the spectral efficiency can be improved by enabling enhanced MIMO techniques.}
However, current communication mechanisms  relying upon  frequency division duplexing (FDD) might not fully support {an} enormous number of devices due to the rate-constrained limited feedback  and the time-consuming scheduling architectures.
%
{In this paper, we propose a user {cooperation-based} limited feedback strategy to support high connection density in massive MTC networks.}
{In the proposed algorithm, two close-in users share {the quantized version of channel information} in order to improve  channel feedback accuracy.}
{The cooperation process is performed without any transmitter interventions (i.e., {in a} grant-free manner) to satisfy the low-latency requirement that is vital for MTC services.}
{Moreover, based on the sum-rate throughput analysis, we develop an adaptive cooperation algorithm {with a view to activating/deactivating the user cooperation mode according to channel and network conditions.}}
%
\end{abstract}


\IEEEpeerreviewmaketitle



\section{Introduction}

\IEEEPARstart{I}{nternet} of things (IoT), {which refers} to the connected future world in which every mobile device and machines are linked to the internet via wireless link, {has received attention from both  academia and industry} in recent years \cite{Ref_Zan14}. {IoT enables {a} wide range of unprecedented services such as autonomous driving, smart home/factory, {and} factory automation, just to name a few \cite{Ref_Lim17}.} Massive connectivity is one of {the} most important requirements {of a} fully connected IoT society \cite{Ref_Boc18,lin20185g}. In accordance with this trend, {the} international telecommunication union (ITU) defined massive machine-type communication (mMTC) {as one representative service category.}\footnote{Three representative service categories include {enhanced} mobile broadband (eMBB), ultra-reliable and low latency communication (uRLLC), and massive machine-type communication (mMTC) \cite{Ref_Boc14}.}
{In mMTC networks,}  data communications may occur between an MTC device and a server or directly between MTC devices \cite{Ref_Tal12}.
It is of {considerable} importance to support high connection density with limited resources because the number of devices is at least two {orders} of magnitude higher than current human-centric communication.

 {From a technological standpoint, enormous number of devices in mMTC networks can be used to exploit full benefit of multiuser MIMO.} {It is essential to have  high-resolution  channel state information (CSI) at the access point (AP) to exploit multiuser diversity gain \cite{Ref_Vis02,Ref_San03}.}
In most  current cellular systems relying upon frequency division duplexing (FDD), the quantized CSI is communicated to the AP via a rate-constrained feedback link \cite{Ref_Lov03,Ref_Lov04,Ref_Son18}.
One challenge of feedback-assisted multiuser systems is that {low-resolution} CSI overrides the multiuser diversity gain because the signal-to-interference-plus-noise ratio (SINR) is limited due to the channel quantization error \cite{Ref_Jin06,Ref_Yoo07}.
In the feedback-assisted multiuser  systems, the rate-constrained feedback mechanism is the biggest obstacle to supporting {a massive number of devices on an MTC network.}

{Antenna combining techniques, e.g.,  \textit{quantization-based combining} (QBC) \cite{Ref_Jin08}, 
has been proposed to obtain  high-resolution CSI.
%
A key feature of the QBC is that receive antennas  are combined to  generate an effective channel that can be quantized accurately.}
%
%
%
Employing more antennas would enhance the quantization performance.
However, direct application of  antenna combining techniques for mMTC {is infeasible} since it is difficult to employ multiple antenna elements due to the strict budget {constraints} on small-scale devices.
%
%
In this paper, we develop a cooperative feedback strategy to obtain an additional spatial dimension for the antenna combining process without imposing an additional burden on mMTC devices.
%

Recently, multiuser systems incorporating user cooperation algorithms have been proposed \cite{Ref_Min17,Ref_Che17}.
{In \cite{Ref_Min17}, the user in the cooperative link helps other adjacent {users by forwarding adjacent users'  information}  while achieving its own quality of service (QoS).
In \cite{Ref_Che17}, the users in the cooperative link exchange their CSI via device-to-device (D2D) communications.
{The  users} can compute {a} more appropriate precoder at the user side because CSI exchange allows users to obtain the global CSI.
However, since the number of users\footnote{{The users can be any kinds of machines, devices, and mobile users.}}} for {the} mMTC network is much larger than that of current human-centric communication, it is not feasible to exchange CSI with all {the} other users. Therefore, it is important to develop solutions with minimal overhead for feedback and/or cooperative links.

%
{{The} aim of this paper is to develop user cooperation strategies for {an} mMTC network allowing only point-to-point CSI exchange between close-in users.
In order to obtain high-resolution CSI {with a minimal burden on} the user cooperation framework, {few} bits are exploited to exchange  CSI.
To the best of our knowledge, {a} user cooperation strategy {designed} to reduce  channel quantization error has been proposed {here} for the first time.
}
%
{The  main contributions of this paper are summarized as follows:}
\
\begin{itemize}
\item \textit{{Cooperative limited feedback:}} {Adjacent users are connected to a cooperation link and these users are considered   one cooperation unit (CU). {{Each user in CU shares the other {user's}}  \textit{local} channel information (i.e., \textit{local} channel direction information (CDI) and channel quality indicator (CQI)).
    {CSI sharing} is only allowed between users in a CU.} 
     Each user generates the \textit{global} channel information required for downlink transmission (i.e., \textit{global} CDI and CQI) using its own channel information and the \textit{local} channel information received from an adjacent user.
    After exchanging each other's \textit{global} CQI, the user having larger \textit{global} CQI is assigned as {the} main user (MU) and the other user is assigned as {an} assistant user (AU). {{The} AU acts as an assistant for MU by allowing  MU to use its receive antennas to construct the \textit{global} CDI more precisely.}     {{The} MU only feeds back the  \textit{global} CSI so that the AP perceives  {the MU} as the sole active user, while {the AU} is transparent to {the AP.}} In the data transmission phase, both {the MU and AU receive}  data information from {the AP} and then {the}  AU forwards the received signal to {the MU.}
    The channel feedback accuracy of {the MU} is improved due to the virtue of exploiting {AU resources.}}
\item \textit{{Automatic role assignment:}} {Identification of the MU and AU} is an important issue since only {the} spectral efficiency of {the MU} can be increased by sacrificing the resources of {the AU.} In the proposed algorithm, {the cooperation process} between users is designed to occur without transmitter intervention (i.e., {a} grant-free environment) through an active decision process. An important issue behind this active decision process is {the} motivation for participating in cooperative communication as {the AU.}\footnote{{One possible option can be {the} social relationship between users \cite{Ref_Che17}. If users have {a} close relationship in the social domain, users can readily help each other by using their own resources {for cooperative feedback.}  {Other possible scenario can be the cooperation between different devices in each user.} {Alternatively,} {an} artificial intelligence (AI)-based and/or game-theoretic approach can be applied in identifying {the} MU and AU and this would be {an interesting future research topic.}}}
    {Under a grant-free environment}, {the AP} regards {the MU as the sole user} and this identification process is transparent to {the AP.}
\item \textit{{Adaptive cooperative feedback:}}
{If cooperative feedback is activated, the number of active users is reduced by half because two users are combined as a single CU to obtain high-resolution CSI.}
Unless {a} massive number of users are active in the mMTC network, the cooperative feedback strategy might not be an effective solution because the multiuser diversity gain is highly limited in a small-user regime. {For this reason, {effective allocation of limited multiuser resources is required to obtain accurate CSI} without loss of multiuser diversity gain.} We analyze the sum-rate throughput of the multiuser MIMO systems relying upon the proposed cooperation algorithm. Based on the analytical studies, {we develop cooperation mode switching criteria to activate/deactivate the cooperation mode according to channel and network conditions.}
\end{itemize}

{In Section II, we briefly introduce a multiuser MIMO system and review a user selection algorithm. In Section III, we present the proposed cooperative feedback algorithm. An adaptive cooperation algorithm is developed based on analytical studies on sum-rate throughput in Section IV. In Section V, we present numerical results to verify the performance of the proposed scheme. We conclude the paper in Section VI.}

{Throughout this paper, $\mathbb{C}$ denotes the field of complex numbers,   $\mathcal{CN}(m,\sigma^2)$ denotes the complex normal distribution with mean $m$ and variance $\sigma^2$, $\mathbf{0}_{a,b}$ is the $a \times b$ all zeros matrix, $\mathbf{1}_{a,b}$ is the $a \times b$ all ones matrix, $\mathbf{I}_{M}$ is the $M \times M$ identity matrix, $\chi^2_{\cdot}$ is the Chi-squared random variable, $\beta(\cdot,\cdot)$ is the Beta-distributed random variable, $\mathbf{B}\big(\cdot,\cdot\big)$ is the Beta function,  $\Gamma(\cdot)$ is the gamma function, $\binom{n}{k}$ is the binomial coefficient, $(z)_{k}$ is the Pochmann symbol, $\lceil~\rceil$ is the ceiling function, $\mathrm{E}[\cdot]$ is the expectation operator, $\mathbbm{1}$ is the indicator function,  $\| \cdot \|_p$ is the $p$-norm, and {$[\ba]_{\ell}$ is the $\ell$-th element of the column vector $\ba$}. 
Also, $\bA_{:,m}$, $\bA^{\dag}$, $\bA^H$, $\mathrm{Tr}(\bA)$,  and $\bA_{a,b}$ denote $m$-th column vector, pseudo-inverse, conjugate transpose, trace, and $(a,b)$-th entry of the matrix $\bA$, respectively.}

\begin{figure*}[!t]
\centering
\subfigure[Conventional multiuser MIMO system.]{\label{fig:system}\includegraphics[width=0.4275\textwidth]{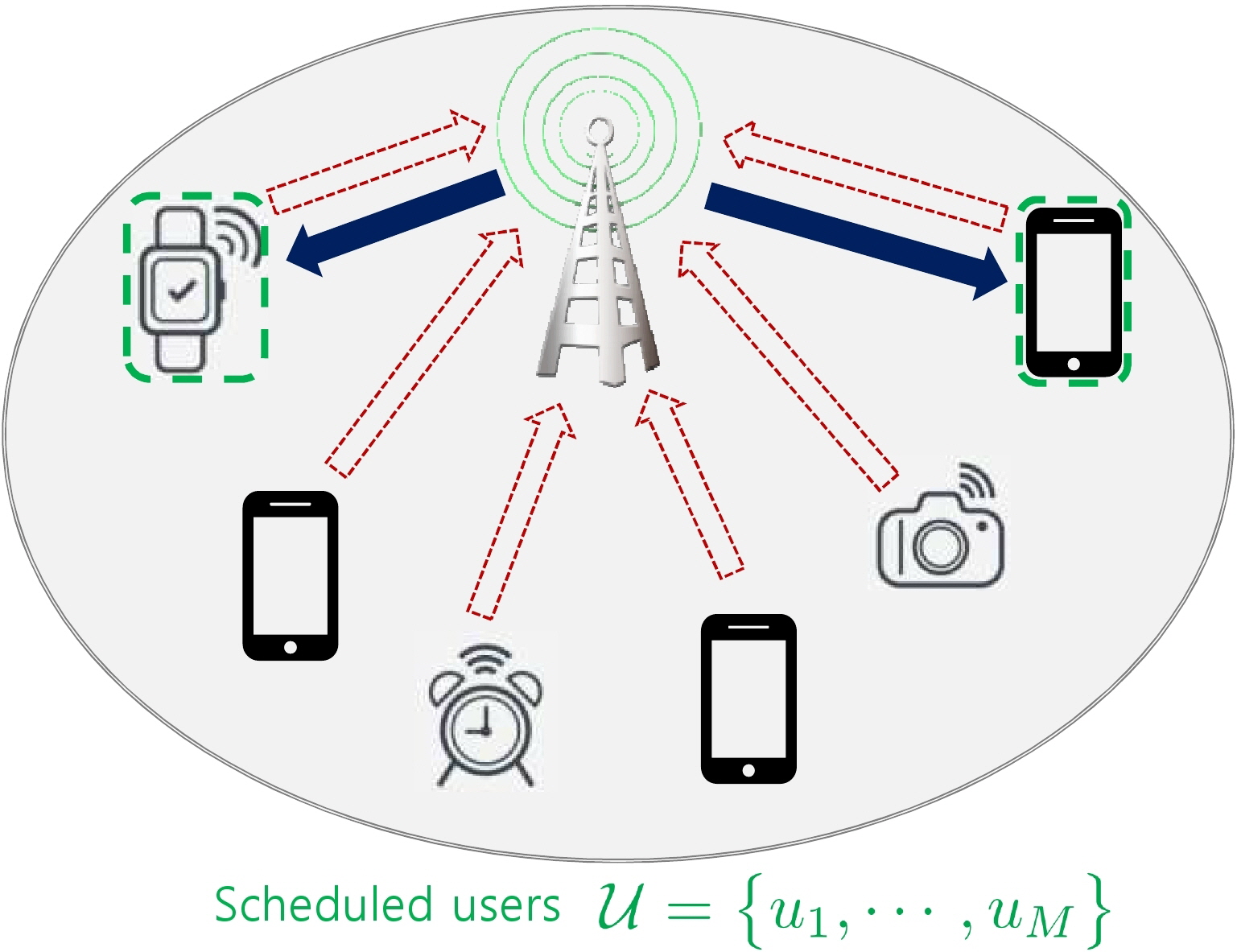}}
\subfigure[Multiuser MIMO with cooperative limited feedback.]{\label{fig:overview}\includegraphics[width=0.56\textwidth]{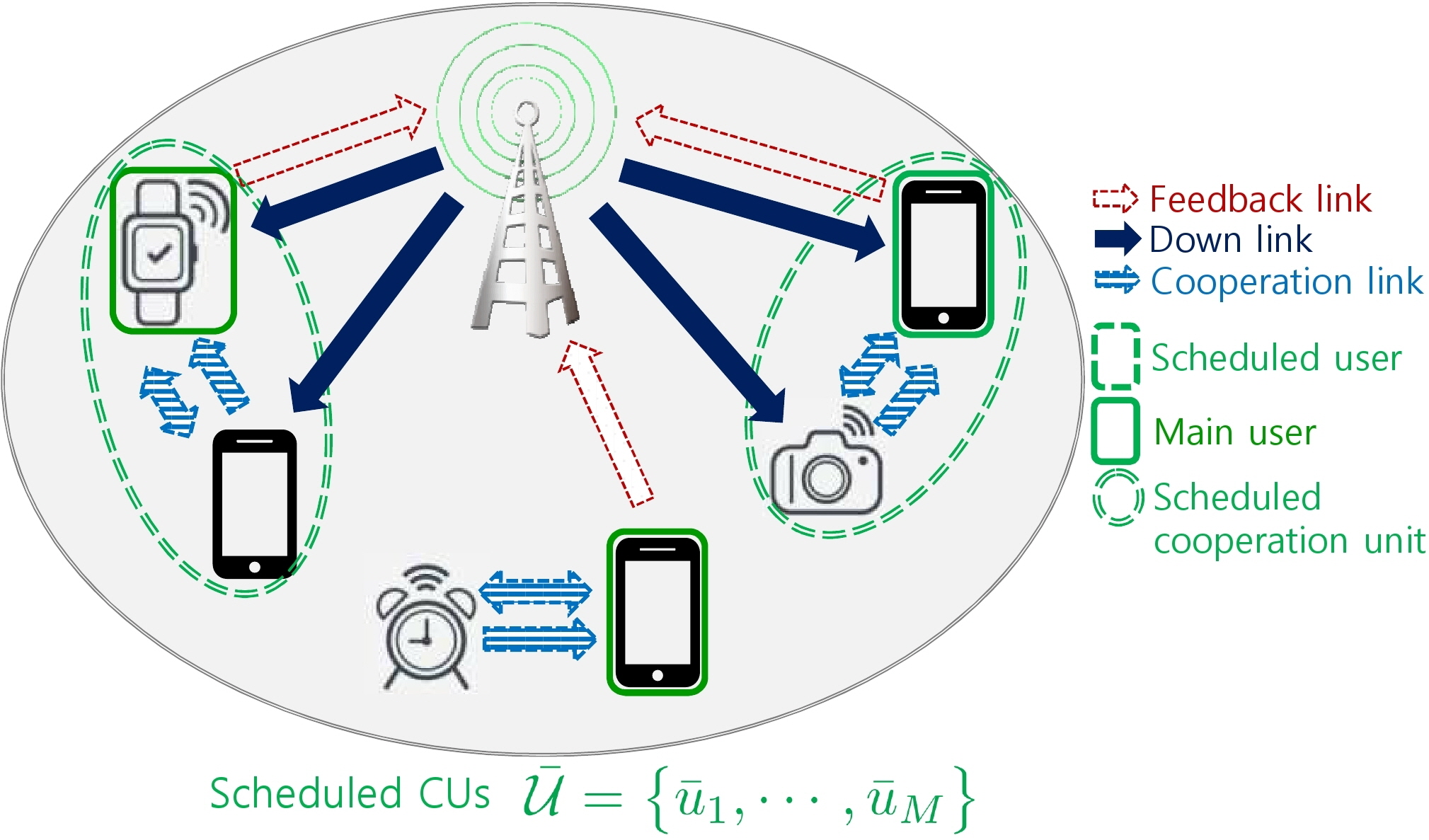}}
\caption{An overview of multiuser MIMO systems.}
\label{fig:system_all}
\end{figure*}


\section{{System Model and Background}}

{We briefly review the FDD-based multiuser MIMO systems.}
We first present the system model and then discuss {\textit{antenna combining}-based} limited feedback. An overview of  conventional multiuser MIMO systems is depicted in Fig. \ref{fig:system}.

\subsection{System Model}
\label{sec:1}
We consider multiuser MIMO systems employing $M$ transmit antennas at the AP and  $N$ receive antennas at each of $K$ users. Assuming a block-fading channel, an input-output expression for the $k$-th user\footnote{{The user index} is subscripted and {the set of user indices} in the network is written by $\mathcal{K}= \{1,\cdots,K \}$.} is  defined as
\begin{align}
 \label{eq:in-out}
{y}_{k}=\bz_{k}^H(\sqrt{\rho}\bH_{k}{\bx}+{\bn}_{k}),
\end{align}
where ${y}_{k} \in \mathbb{C}$ is the received baseband signal, $\rho$ is the signal-to-noise ratio (SNR), $\bz_{k} \in \mathbb{C}^N$  is the unit-norm  combiner,
\begin{align}
\label{eq:channel_MIMO}
\bH_{k} \doteq \big[{\bh}^1_{k},\cdots,{\bh}^N_{k}\big]^{H} \in \mathbb{C}^{N \times M}
\end{align}
is the MIMO channel matrix,  {$\bh^{n}_{k} \in \mathbb{C}^M$ is the channel vector between the AP and the $n$-th receive antenna consisting of independent and identically distributed (i.i.d.) entries following $\mathcal{CN}(0,1)$,} $\bx \in \mathbb{C}^M$ is the transmit signal vector (with the power constraint $\mathrm{E} [ \| \bx\|_2^2  ] \leq 1$), and {${\bn_{k}} \in \mathbb{C}^N$ is the additive noise vector with entries following $\mathcal{CN}(0,1)$.}

We consider a single layer data transmission for each user.
The transmit signal vector is rewritten as $\bx \doteq \bF\bs$, where
\begin{align*}
\bF=\frac{1}{\sqrt{M}}[{\bff}_{1},\cdots,{\bff}_{M}] \in \mathbb{C}^{M \times M}
\end{align*}
is the precoder and  ${\bs}=[s_{1},\cdots,s_{M}]^{T} \in \mathbb{C}^M$ is the transmit symbol vector. Note that $\bff_{m} \in \mathbb{C}^M$ and $s_{m} \in \mathbb{C}$ denote the transmit beamformer and the data stream for the $m$-th scheduled user with the power constraints $\|\bff_m\|_2^2=1$ and $\mathrm{E} [|s_m|^2] \leq 1$.

%
In FDD-based cellular systems, an AP obtains channel information through receiver feedback from each user.
%
In feedback-assisted MIMO architectures, channel vectors are quantized   using the predefined   codebook
\begin{align}
\label{eq:global_cbk}
\mathcal{C} \doteq \{\bc^1 ,\cdots, \bc^{Q} \}.
\end{align}
where $Q=2^B$ is the number of codewords in the \textit{global} codebook.
To facilitate the multiuser signaling framework, the quantized channel information is fed back to the AP via a rate-constrained $B=\lceil\log_2{Q}\rceil$-bit feedback link.

We employ the opportunistic random beamforming approach that utilizes a set of unitary matrices {to construct} the \textit{global} codebook \cite{Ref_Vis02,Ref_Chu03}.
{Similar} to the  LTE-Advanced codebooks in \cite{Ref_3GPP103378,Ref_3GPP105011}, exploiting more sets of unitary matrices (i.e., oversampled discrete Fourier transform (DFT) codebook) enables the AP to obtain  high-resolution CSI.
However, an  ultra low-latency requirement for MTC services restricts the use of {large  codebooks.}
{Assuming an intensely rate-constrained feedback link, we consider  $M$ codewords for  channel feedback and beamforming, meaning that the number of codewords equals to the number of transmit antennas.
If only $M$ codewords are allowed for CSI quantization, it would be optimal to consider a set of  orthonormal vectors for random beamforming  \cite{Ref_Kou08}.}
We use a single unitary matrix $\bC \doteq [\bc^1,\cdots, \bc^M]$  for defining  codewords according to $\bc^m\doteq \bC_{:,m}$.

\subsection{{QBC-based Limited Feedback for Multiuser MIMO}}
\label{sec:review}

%
%
One major issue of the feedback architecture using $M$ codewords is that
the channel quantization performance is expected to be poor in a single receive antenna scenario.
When  multiple receive antennas are available, antenna combining techniques can be applied to enhance the channel quantization performance \cite{Ref_Jin08,Ref_Tri08}.
%
In this paper, we consider {a QBC-based} antenna combining algorithm \cite{Ref_Jin08}.
%
The objective of the QBC algorithm is to compute an effective channel vector that can be quantized accurately using a small-sized codebook.
For a given target codeword $\bc^m$, the receive combiner $\bar{\bz}_{k|m}$ is computed such that a cross-correlation between the effective channel
\begin{align}
\label{eq:eff_MISO}
\bar{\bh}_{k|m}^{\mathrm{eff}}=\bH_{k}^H\bar{\bz}_{k|m}
\end{align}
and the target codeword  is maximized. 
{As discussed in \cite{Ref_Jin08}, the effective channel that maximizes the cross-correlation is obtained by projecting  $\bc^m$ onto the channel subspace such that
\begin{align}
\label{eq:proj_codeword}
 {\bc}^{\mathrm{proj},m}   =\frac{[\bq^1,\cdots,\bq^N][\bq^1,\cdots,\bq^N]^H\bc^m}{\big\| [\bq^1,\cdots,\bq^N][\bq^1,\cdots,\bq^N]^H\bc^m \big\|_2} \in \mathbb{C}^{M},
\end{align}
where $\bq^n$ is the orthonormal basis that spans  $\bH_{k}$. Using the QBC algorithm, the receive combiner $\bar{\bu}_{k|m}$, which satisfies $\bH_{k}^H\bar{\bu}_{k|m}={\bc}^{\mathrm{proj},m}$, can be computed by multiplying the pseudo-inverse of the channel matrix such as}
\begin{align*}
(\bH_{k}\bH_{k}^H)^{-1}\bH_{k}\bH_{k}^H\bar{\bu}_{k|m}=(\bH_{k}\bH_{k}^H)^{-1}\bH_{k}{\bc}^{\mathrm{proj},m}.
\end{align*}
Finally, the unit-norm  combiner is computed according to
\begin{align}
\label{eq:combiner_nocoop}
\bar{\bz}_{k|m} \doteq \frac{\bar{\bu}_{k|m}}{\|\bar{\bu}_{k|m} \|_2} = \frac{ (\bH_{k}\bH_{k}^H)^{-1}\bH_{k}{\bc}^{\mathrm{proj},m}  }{\|(\bH_{k}\bH_{k}^H)^{-1}\bH_{k}{\bc}^{\mathrm{proj},m}   \|_2} \in \mathbb{C}^{N}.
\end{align}

%
Assuming the user $k$ is scheduled to use {the} $m$-th   beamformer (codeword), the received signal is written by\
\begin{align}
\label{eq:re-input-output}
&{y}_{k|m} =  \sqrt{\rho} \bar{\bz}_{k|m}^H \bH_{k}\bigg(\frac{[{\bc}^{1},\cdots,{\bc}^{M}]}{\sqrt{M}} \bigg) \bs +\bar{\bz}_{k|m}^H \bn_{k}
\\
\nonumber
& =\sqrt{\frac{\rho}{M}}  \Bigg(  (\bar{\bh}_{k|m}^{\mathrm{eff}})^H  \bc^{{m}}s_{m}+ \sum_{\ell=1,\ell \ne m}^M (\bar{\bh}_{k|m}^{\mathrm{eff}})^H \bc^{\ell}s_{\ell} \Bigg)+ {\bar{n}_{k|m}},
\end{align}
because the $m$-th  beamformer is identical to the $m$-th codeword such that $\bff_m=\bc^m$ in our random beamforming architecture.
Notice that  $\bar{n}_{k|m} \doteq \bar{\bz}_{k|m}^H{\bn_{k}}  \sim \mathcal{CN}(0,1)$ denotes the combined noise. {The SINR of the $k$-th user is then defined by\footnote{{In our beamforming scenario exploiting a single unitary matrix, the SINR can be computed at the receiver side because the $k$-th user knows all the transmit beamformers, i.e.,  $\bc^{\ell},~\ell \ne k$, assigned  for other  users. The SINR can be regarded as CQI that quantifies the quality of  each transmission layer.}}}
\begin{align}
\label{eq:SINR_initial}
\bar{\gamma}_{k|m}\doteq \frac{|(\bar{\bh}_{k|m}^{\mathrm{eff}})^H  \bc^m|^2}{ \frac{M}{\rho} +   \sum_{\ell=1,\ell \ne m}^M| (\bar{\bh}_{k|m}^{\mathrm{eff}})^H \bc^{\ell} |^2}.
\end{align}
%

From among $M$  beamformers  $\{ \bc^1,\cdots,\bc^M\}$, each user (e.g., $k$-th user) selects a single  beamformer
\begin{align}
\label{eq:set_nocoop}
 \bar{\bv}_{k} \doteq \bc^{\hat{m}}
\end{align}
{that maximizes the SINR   according to $\bar{\gamma}_{k}=\bar{\gamma}_{k|\hat{m}}$, where the index of the selected codeword is $\hat{m} \doteq \argmax_{m}\bar{\gamma}_{k|m}$, and the selected combiner is\footnote{{For the sake of simplicity, the index of the selected codeword is dropped for the rest of the sections.}} $\bar{\bz}_{k} = \bar{\bz}_{k|\hat{m}}$.}
{We call the selected beamformer $\bar{\bv}_{k}$   \textit{global} CDI  and the selected SINR $\bar{\gamma}_{k}$   \textit{global} CQI.}
In this paper,  we focus on quantizing the CDI and refer to \cite{Ref_Son12} and the references therein for quantizing the CQI. We assume that  the index of the quantized CDI is fed back  via an error-free $B$-bit feedback link  and the unquantized CQI can be communicated to the AP.

Finally, we refer to the  user selection algorithm in \cite{Ref_Son12} to schedule/select $M$ users from among $K \gg M$ users such that
\begin{align*}
\mathcal{U}=\big\{u^1,\cdots ,u^M \big\},
\end{align*}
where $u^{m}$ denotes the scheduled user exploiting the $m$-th codeword $\bc^m$.
The $m$-th scheduled user is given by
\begin{align*}
u^{m} \doteq \argmax_{k \in \mathcal{K}^m} \gamma_{k},
\end{align*}
where $\mathcal{K}^m$ denotes the set of indices of users who choose $\bc^m$ as their \textit{global} CDI.
{It should be noted that the multiuser diversity gain from user selection plays a significant role in improving the sum-rate throughput and it grows like $M \log_2 (\log K)$ under the perfect CSI assumption at the AP \cite{Ref_Vis02,Ref_Sha07}.}

{Despite the advantage, it has some obstacles that hinder the direct application of conventional multiuser systems to mMTC.
First, the channel quantization error overrides the multiuser diversity gain.
In feedback-assisted FDD architectures, multiuser systems become interference-limited due to   unsuppressed  quantization error.
%
%
%
%
The sum-rate  is thus upper bounded {even though} the SNR goes to infinity  \cite{Ref_Jin06,Ref_Yoo07}.
Second,  a large number of  devices imposes a heavy burden on the initial access architecture of cellular networks.
In 5G new radio (NR), there are a limited number of  physical-layer cell identities \cite{lin20185g}.
Since there is no specific collision avoidance procedure, a massive number of access attempts can cause severe congestions that accompany the increase in transmission latency \cite{Ref_Boc18}.
Even assuming congestion-free scenarios, the sum-rate growth in a large-user regime $K \gg M$ has slowed because the sum-rate grows in a double-logarithmical fashion \cite{Ref_Sha05,Ref_Sha07}.}


\section{Cooperative Limited Feedback Architecture}

%

{One of the key challenges in developing feedback-assisted multiuser systems for mMTC is to obtain accurate CSI at the AP while overcoming the following restrictions:
\begin{itemize}
\item Ultra low-latency requirement for mMTC restricts the use of large  codebook and this {make it} difficult to achieve robust channel quantization performance.
\item Considering the strict budget constraints of small-scale devices, it is not practical to employ a multitude of antenna elements  for the antenna combining.
\end{itemize}
In QBC, the resolution of an effective channel  increases as the spatial dimensions of a channel matrix for an antenna combining increases \cite{Ref_Jin08}.
In this paper, we propose user cooperation strategies   to obtain   high-resolution CSI  while limiting the number of antenna elements at the receiver.
The main feature of the proposed  algorithm is that users in a CU employ an additional spatial dimension for the antenna combining  by allowing  a limited
amount of CSI exchange.
The key difference between the conventional system in Fig. \ref{fig:system} and the proposed system in Fig. \ref{fig:overview} is the existence of the cooperation link.
We assume that the quantized version of CSI can be exchanged between close-in users by using the Wi-Fi peer-to-peer sidelink, as presented in  \cite{Ref_Shi10}.}


\begin{figure*}[!t]
\centering
\subfigure{\includegraphics[width=0.8575\textwidth]{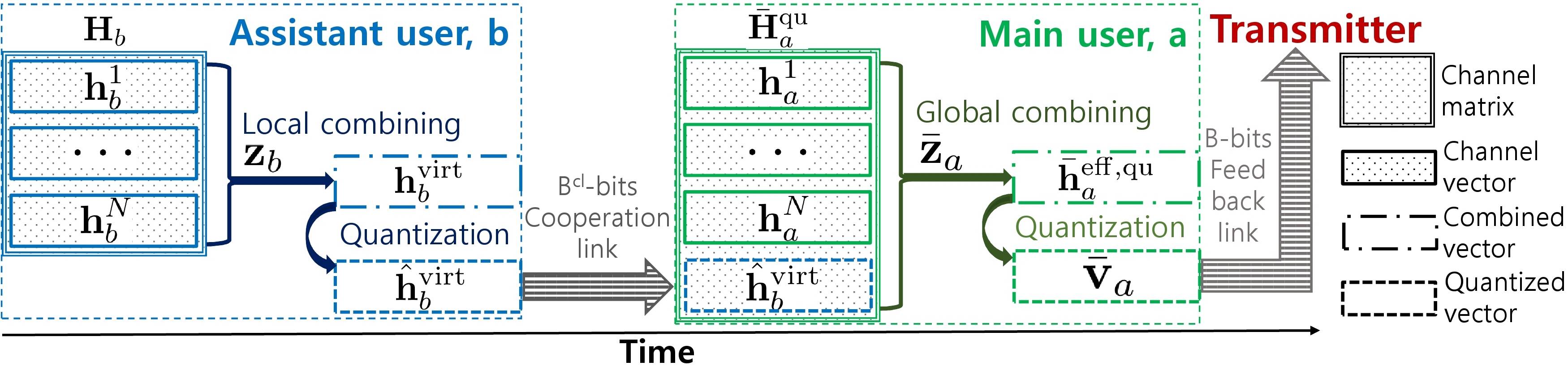}}
\caption{An overview of \textit{global} and \textit{local} combining processes.}
\label{fig:combine}
\end{figure*}

 {Before presenting the cooperative limited feedback algorithm, we pause to provide supporting details behind the variables in the proposed cooperative feedback architecture:}
\begin{itemize}
\item The term \textit{global} is used to designate the variables and  signal processing operations within CU, while the term \textit{local} is used to  designate the variables and  signal processing operations within AU.
\item The bar symbol $\bar{\cdot}$ will be used to highlight {variables} corresponding  to the \textit{\textit{global}} signal processing operation.
\item The tilde symbol $\tilde{\cdot}$ will be used to highlight {variables} corresponding to the downlink transmission.
\end{itemize}



{We present the details of the cooperative feedback algorithm based on the assumption that two close-in users have already been combined as a single CU\footnote{Developing {a} user grouping algorithm for holding two users together to form   CU and/or considering a coalition of more than two users in {a} CU would be {an} interesting future research topic.} according to $ \{a,b\}$.} 

\subsubsection{\textbf{{Local CSI acquisition}}}
\label{sec:local_comb}
{An aim of this  step is to compute   \textit{local} CSI that will be used to increase spatial dimensions of the partner's channel matrix. The quantized version of \textit{local} CSI is  transferred to its partner via a cooperation link.}

{One viable approach to achieve reduction of both \textit{local} CSI quantization error and cooperation link overhead is to share only a single effective channel vector that is combined based on the QBC algorithm \cite{Ref_Jin08}.
The combined channel  is quantized   using the  random vector quantization (RVQ) codebook
\begin{align}
\label{eq:local_cbk}
\mathcal{D} \doteq \{\bd^1,\cdots, \bd^{{Q}^{\mathrm{cl}}} \},
\end{align}
which consists of ${Q}^{\mathrm{cl}} = 2^{{B}^{\mathrm{cl}}}$ codewords.
%
%
The distance of the cooperation link  is much shorter than that of the feedback link. The cooperation link  would be subject to less stringent overhead constraints compared to that for the feedback link such that ${B}^{\mathrm{cl}} \gg B$.}

{First, user  {$k \in \{a,b\}$} in the CU  computes a virtual  vector
\begin{align}
\label{eq:virtual_vector}
{\bh}_{k|q}^{\mathrm{virt}}  \doteq \bH_k^H \bz_{k|q}
\end{align}
{that can be quantized more accurately using a target codeword $\bd^q  \in \mathcal{D}$, where the  \textit{local} combiner $\bz_{k|q} \in \mathbb{C}^{N}$ and the projected codeword ${\bd}^{\mathrm{proj},q} \in \mathbb{C}^{M}$, needed for the \textit{local} channel combining,  are  computed   using similar  method in (\ref{eq:proj_codeword}) and (\ref{eq:combiner_nocoop}).}} 


{Second, each user selects a  \textit{local} codeword that maximizes {the} cross-correlation of the normalized virtual channel vector  and the codeword, i.e.,
\begin{align}
\label{eq:selected beamforming_gain}
\cos^2\phi_{k|q}=\big| (\bd^q)^H {\bh}_{k|q}^{\mathrm{virt}} /\| {\bh}_{k|q}^{\mathrm{virt}} \|_2 \big|^2,
\end{align}
{where $\phi_{k|q}$ is the difference in angle.} {We call the selected codeword  \textit{local}  CDI and its corresponding cross-correlation coefficient   \textit{local}  CQI.}
The   \textit{local} CDI and CQI  are  given by
\begin{align}
\label{eq:sel_local}
\big({\bv}_k, \tau_k \big)  &\doteq \big(\bd^{\hat{q}},\|{\bh}_{k|\hat{q}}^{\mathrm{virt}}\|_2 \cos \phi_{k|\hat{q}} \big),
\end{align}
where the index of the selected codeword is
\begin{align*}
\hat{q} =\argmax_{q \in \{1,\cdots,{{Q}^{\mathrm{cl}}}\} } \cos^2\phi_{k|q}.
\end{align*}
{Under the assumption the index of {the} selected codeword is dropped, the selected  \textit{local} combiner, the virtual channel vector, and the difference in angle can be rewritten as}
\begin{align}
\label{eq:unquan_local}
\big(\bz_k, {\bh}_{k}^{\mathrm{virt}} ,\phi_{k} \big) \doteq \big(\bz_{k|\hat{q}},{\bh}_{k|\hat{q}}^{\mathrm{virt}},\phi_{k|\hat{q}} \big).
\end{align}
The quantized virtual channel vector is then  defined with the \textit{local} CDI and CQI according to}
\begin{align}
\label{eq:quan_virtual}
\hat{\bh}_{k}^{\mathrm{virt}} \doteq \tau_k \bv_k=\|{\bh}_{k}^{\mathrm{virt}}\|_2 \cos \phi_{k}\bv_k.
\end{align}

{Finally, users in the CU  exchange the \textit{local} CDI and CQI, i.e., quantized virtual channel vector $\hat{\bh}_{k}^{\mathrm{virt}}$ in (\ref{eq:quan_virtual}), with its cooperation partner via a ${B}^{\mathrm{cl}}$-bits cooperation link.
{The \textit{local} CDI and CQI will be included in a \textit{global} channel matrix of its cooperation partner.
The \textit{local}  combining and  quantization processes are depicted in the left side of Fig. \ref{fig:combine}.}}

\subsubsection{\textbf{{Global CSI acquisition $\&$ Role assignment}}}
\label{sec:global_comb}
 {An aim of this  step is to compute \textit{global} channel information and to assign the role of MU and AU. The \textit{global} CSI  will be fed back to the AP.}

{First, each user   constructs a \textit{global} channel matrix
\begin{align}
\label{eq:virtual_matrix}
{\bar{\bH}^{\mathrm{qu}}}_k \doteq \bigg[\begin{array}{c}
                \bH_k \\
                (\hat{\bh}_{{k}^{\mathrm{c}}}^{\mathrm{virt}})^H
\end{array}\bigg] \in \mathbb{C}^{(N+1) \times M},
\end{align}
which includes one's own channel matrix $\bH_k$  and the quantized virtual channel vector $\hat{\bh}_{{k}^{\mathrm{c}}}^{\mathrm{virt}}$ from a cooperation partner, where ${k}^{c} \in \{a,b \} \setminus \{k\}$.} 
{Each user regards the  virtual channel vector as an additional channel vector between the AP and the virtual antenna element at the receiver.}
{Assuming oneself is selected as MU, each user computes {the} \textit{global} effective channel vector
\begin{align}
\label{eq:effective_vector}
\bar{\bh}_{k|m}^{\mathrm{eff,qu}} =({\bar{\bH}^{\mathrm{qu}}}_{k})^H \bar{\bz}_{k|m}
\end{align}
{that can be quantized   accurately with a target codeword $\bc^m \in \mathcal{C}$.}
%
The \textit{global} combiner $\bar{\bz}_{k|m} \in \mathbb{C}^{N+1}$ is  computed using the combining method in (\ref{eq:proj_codeword}) and (\ref{eq:combiner_nocoop}).}
The difference with the combining process in (\ref{eq:virtual_vector}) is that the rank of the channel matrix and the dimension of  the combiner are increased to $N+1$.

Second, each user selects a single \textit{global} codeword that maximizes the SINR
\begin{align*}
\bar{\gamma}_{k|m}&\doteq \frac{|(\bar{\bh}_{k|m}^{\mathrm{eff,qu}})^H \bc^m|^2}{\frac{M}{\rho}+ \sum_{\ell=1,\ell \ne m}^M | (\bar{\bh}_{k|m}^{\mathrm{eff,qu}})^H   \bc^\ell |^2}.
\end{align*}
We call the selected codeword   \textit{global}  CDI and its corresponding SINR  \textit{global}   CQI.
The \textit{global} CDI  and CQI are
\begin{align}
\label{eq:sel_global}
\big(\bar{\bv}_k  , \bar{\gamma}_{k} \big) \doteq \big( \bc^{\hat{m}} ,   \bar{\gamma}_{k|\hat{m}}\big),
\end{align}
where the index of the selected codeword is $\hat{m} \doteq \argmax_{m}\bar{\gamma}_{k|m}$. Assuming  the index of {the} selected codeword is dropped, the selected  \textit{global} combiner and the effective channel vector are rewritten according to
\begin{align}
\label{eq:unquan_global}
\big(\bar{\bz}_k, \bar{\bh}_{k}^{\mathrm{eff,qu}} \big) \doteq \big(\bar{\bz}_{k|\hat{m}}, \bar{\bh}_{k|\hat{m}}^{\mathrm{eff,qu}} \big).
\end{align}

Third, users  in {the} CU  exchange  their \textit{global} CQIs with {cooperating} users via a cooperation link {to determine who's best for maximizing the data-rate throughput.}
The user having a larger CQI is assigned {to} MU  and the unselected user is assigned {to} AU.
{{The  role assignment  are made at the beginning of the transmission frame and will continue for the duration of the channel coherence time.}
We  assume that the $a$-th (odd number indexed) user is assigned {to} MU and the $b$-th (even number indexed $b=a+1$) user is assigned {to} AU. The set of indices of MUs is then  written by $\mathcal{\bar{K}}=\{1,3,\cdots, K-1\}$.}
%

{Finally, MU only transmits the \textit{global} CSI to the AP via a  feedback link.
%
%
{The number of active users in the network is thus reduced by half  $\big| \mathcal{\bar{K}}\big|=K/2$.}}
{The \textit{global} combining  and   quantization processes are depicted in the right side of  Fig. \ref{fig:combine}.}\

\subsubsection{\textbf{{User scheduling}}} {After collecting \textit{global} CDI and CQI from $K/2$ MUs, the AP schedules $M$ MUs such that
\begin{align*}
\bar{\mathcal{U}} = \big\{ \bar{u}^1,\cdots, \bar{u}^M \big\},
\end{align*}
where $\bar{u}^m$ denotes the scheduled MU exploiting the $m$-th codeword $\bc^m$.
The $m$-th scheduled MU is given by
\begin{align*}
\bar{u}^m = \argmax_{k \in \bar{\mathcal{U}}^m} \bar{\gamma}_{k},
\end{align*}
where  $\bar{\mathcal{U}}^m$ denotes the set of indices of MUs who choose $\bc^m$  as their  \textit{global} CDI.}

\begin{algorithm}
  \caption{Cooperative feedback algorithm}
  \label{Al:01}
  \begin{algorithmic}
\State \textbf{Step 1) {Local CSI acquisition}}
\State ~1:~~{Compute \textit{local} combiner and virtual channel vector}
\State ~~~~~$\bz_{k|q}=\frac{(\bH_{k}\bH_{k}^H)^{-1}\bH_{k} {\bd}^{\mathrm{proj},q} }{ \| (\bH_{k}\bH_{k}^H)^{-1}\bH_{k} {\bd}^{\mathrm{proj},q}  \|_2},~{\bh}_{k|q}^{\mathrm{virt}}=\bH_{k}^H\bz_{k|q}$
\State ~2:~~{Select \textit{local} CDI and CQI} $({\bv}_{k},\tau_{k})$
\State ~3:~~\textit{Exchange \textit{local} CDI and CQI   $\hat{\bh}_{k}^{\mathrm{virt}} = \tau_k \bv_k$}
\State \textbf{Step 2) {Global CSI acquisition $\&$ Role assignment}}
\State ~4:~~{Construct \textit{global} channel matrix} {$\bar{\bH}^{\mathrm{qu}}_k=[\bH_k^H, \hat{\bh}_{{k}^{\mathrm{c}}}^{\mathrm{virt}}]^H$}
\State ~5:~~{Compute \textit{global} combiner and effective channel vector}
\State ~~~~~{$\bar{\bz}_{k|m}=\frac{(\bar{\bH}^{\mathrm{qu}}_k (\bar{\bH}^{\mathrm{qu}}_k)^H)^{-1} \bar{\bH}^{\mathrm{qu}}_k  {\bc}^{\mathrm{proj},m}  }{\| (\bar{\bH}^{\mathrm{qu}}_k (\bar{\bH}^{\mathrm{qu}}_k)^H)^{-1} \bar{\bH}^{\mathrm{qu}}_k {\bc}^{\mathrm{proj},m}   \|_2},$}
\State ~~~~~{$\bar{\bh}_{k|m}^{\mathrm{eff,qu}}=(\bar{\bH}^{\mathrm{qu}}_k)^H \bar{\bz}_{k|m}$}
\State ~6:~~{Select \textit{global} CDI and CQI} $(\bar{\bv}_k, \bar{\gamma}_{k} )$
\State ~7:~~\textit{Exchange \textit{global} CQI}
\State ~8:~~{Assign MU having a larger \textit{global}  CQI}
\State ~9:~~\textit{MU reports \textit{global} CDI and CQI to the AP}
\State \textbf{Step 3) {User scheduling}}
\State ~10:~{Schedule $M$ MUs} $\bar{\mathcal{U}} = \{ \bar{u}^1,\cdots, \bar{u}^M \}$
\State \textbf{Step 4) {Post-signal processing}}
\State ~11:~{Save received signals} {$\by_{k}$}
\State ~12:~{AU combines received signals} {$y_b=\bz_b^H\by_b$}
\State ~13:~\textit{AU reports $y_b$ to MU}
\State ~14:~{MU obtains virtual received signals} {$\bar{\by}_a=[\by_a^T,y_b^T]^T$}
\State ~15:~{MU combines received signals} {$\bar{y}_{a|m}=\bar{\bz}_{a|m}^H\bar{\by}_a$}
  \end{algorithmic}
\end{algorithm}

\subsubsection{\textbf{{Post-signal processing}}} An aim of this  step is to decode the received signals\footnote{{A memory in the receiver allows retention of received signals. The post-signal processing has no effect on a variation of the sum-rates because this process is conducted using the received signals stored in memory.}}
\begin{align}
\label{eq:received}
\by_{k}=\sqrt{\rho}\bH_{k}\bx+\bn_{k} \in \mathbb{C}^N,~k \in \{a,b \}.
\end{align}
%
{First, AU  combines the received signal  with the \textit{local} combiner $\bz_b$ such as
\begin{align}
\label{eq:signal_AU}
y_b &= \sqrt{\rho}\bz_b^H\bH_{b}\bx+\bz_b^H\bn_b =\sqrt{\rho}({\bh}_{b}^{\mathrm{virt}})^H\bx+n_b \in \mathbb{C},
\end{align}
where ${\bh}_{b}^{\mathrm{virt}}$ is the unquantized virtual channel vector in (\ref{eq:unquan_local}).}
Second, the  combined signal  $y_{b}$  is  passed from AU to  MU. Then, {the  MU} constructs the  \textit{global} signal vector
\begin{align}
\nonumber
\bar{\by}_{a}=
\bigg[\begin{array}{c}
                \by_{a} \\
                y_{b}
\end{array}\bigg]
&=
\sqrt{\rho}
\bigg[\begin{array}{c}
                \bH_{a} \\
                ({\bh}_{b}^{\mathrm{virt}})^H
\end{array}\bigg]
\bx
+
\bigg[\begin{array}{c}
                \bn_a \\
                n_b
\end{array}\bigg]
\\
\label{eq:signal_MU}
&=\sqrt{\rho}\bar{\bH}_a\bx +\bar{\bn}_{a}
\in \mathbb{C}^{N+1},
\end{align}
{where the \textit{global} channel matrix corresponding to downlink transmission (downlink channel matrix) is defined by
\begin{align}
\label{eq:virtual_matrix_trans}
\bar{\bH}_a \doteq \bigg[\begin{array}{c}
                \bH_a \\
               ({\bh}_{b}^{\mathrm{virt}})^H
\end{array}\bigg] \in \mathbb{C}^{(N+1) \times M},
\end{align}
and the \textit{global} noise vector is  $\bar{\bn}_{a} \doteq [ \bn_{a}^T, ~n_b^T]^T \in \mathbb{C}^{N+1}$.}
{Finally,  {the MU} combines the \textit{global} signal vector with the  \textit{global} combiner   according to
\begin{align}
\label{eq:re_combine}
\bar{y}_{a|m}&=\bar{\bz}_{a|m}^H\bar{\by}_{a} \in \mathbb{C}.
\end{align}
%
{The detailed steps of the proposed algorithm are {presented}  in Algorithm \ref{Al:01} and  important variables  are written in Table \ref{tab:summary}.}

\def\RowStrut{\raisebox{-.33\height}{\hbox to0pt{\rule{0pt}{14pt}}}}
\begin{table}
\caption{Summary of important variables}
\centering
\begin{tabular}{|l|l|l|c|}
  \hline
     \RowStrut  \textit{Local}  &  Assistant user  & $b \in \{2,4,\cdots,K \}$  &      \\ \cline{2-4}
     \RowStrut    &Channel matrix  & $\bH_b = \big[{\bh}^1_b,\cdots,{\bh}^N_b \big]^{H}$ & (\ref{eq:channel_MIMO})     \\ \cline{2-4}
       \RowStrut  &Codeword &$\bd^q \in \mathcal{D}$ & (\ref{eq:local_cbk})     \\ \cline{2-4}
       \RowStrut  &Codebook size &$\big|\mathcal{D}\big|=2^{{B}^{\mathrm{cl}}}={Q}^{\mathrm{cl}}$ & (\ref{eq:local_cbk})     \\ \cline{2-4}
            \RowStrut &Combiner & $\bz_{b|q} $ &     (\ref{eq:combiner_nocoop})  \\ \cline{2-4}
     \RowStrut  &Virtual vector & ${\bh}_{b|q}^{\mathrm{virt}}=\bH_b^H \bz_{b|q}$  &   (\ref{eq:virtual_vector})  \\ \cline{2-4}
      \RowStrut &CDI and CQI & $({\bv}_b,\tau_b)$  &  (\ref{eq:sel_local}) \\ \cline{2-4}
    \RowStrut &Quantized  vector & $\hat{\bh}_{b}^{\mathrm{virt}} = \tau_b\bv_b$  &  (\ref{eq:quan_virtual}) \\ \hline

    \RowStrut \textit{Global}  &  Main user  & $a \in \{1,3,\cdots,K-1 \}$  &      \\ \cline{2-4}
    \RowStrut &Channel matrix  & $\bar{\bH}^{\mathrm{qu}}_a=\big[{\bH}_a^H, \hat{\bh}_{b}^{\mathrm{virt}} \big]^H $ & (\ref{eq:virtual_matrix})     \\ \cline{2-4}
    \RowStrut    &Codeword&$ \bc^m \in \mathcal{C}$ & (\ref{eq:global_cbk})    \\ \cline{2-4}
     \RowStrut  &Codebook size &$\big|\mathcal{C}\big|=2^{B}=M$ & (\ref{eq:global_cbk})   \\ \cline{2-4}
         \RowStrut   &Combiner & $\bar{\bz}_{a|m}$ &  (\ref{eq:combiner_nocoop})    \\ \cline{2-4}
     \RowStrut   &Effective  vector & $\bar{\bh}_{a|m}^{\mathrm{eff,qu}} =(\bar{\bH}^{\mathrm{qu}}_a)^H \bar{\bz}_{a|m} $ &    (\ref{eq:effective_vector})  \\ \cline{2-4}
    \RowStrut    &CDI and CQI& $(\bar{\bv}_a,\bar{\gamma}_{a}) $  &  (\ref{eq:sel_global})  \\   \hline
\end{tabular}
\label{tab:summary}
\end{table}


%

\section{Adaptive Cooperation Algorithm}
\label{sec:adaptive}

{{The proposed} cooperative feedback algorithm exploits {some  multiuser resources  to improve}  channel quantization performance.
{High-resolution CSI} can be obtained  because more antenna elements (spatial dimensions) are used for   \textit{global} antenna combining.
However, the proposed approach would restrict {options that could be used to improve} a network throughput due to the following reasons:
%
%
First, the squared norm of the \textit{global} effective channel vector decreases as the number of antennas used for a combining process increases \cite[Lemma 3]{Ref_Jin08}.
Second, the sum-rate grows like $M \log_2 (\log K/2)$ because user candidates  are reduced by half.}
%

{In this section, we develop an analytical framework {weighing} the pros and cons of the proposed cooperative feedback algorithm. Based on the analytical framework, an adaptive cooperation algorithm is proposed in order to activate/deactivate the proposed cooperation strategy according to  channel and  network conditions.}

\subsection{Loss in Local Channel Quantization}
\label{sec:quantization_error}

Before investigating the received signal of MU, we pause to analyze the channel quantization error induced in the process of \textit{local} combining.
{As discussed in Section \ref{sec:review},
the channel quantization error between the normalized virtual channel vector $\frac{{\bh}_{b|q}^{\mathrm{virt}}}{\| {\bh}_{b|q}^{\mathrm{virt}} \|_2} $ and the target codeword $\bd^q$  is quantified by}
\begin{align}
\label{eq:quan_error}
\mathrm{S} \doteq \sin^{2}\phi_{b|q} &= 1-\big| (\bd^q)^H {\bh}_{b|q}^{\mathrm{virt}} /\| {\bh}_{b|q}^{\mathrm{virt}} \|_2 \big|^2,
\end{align}
where $\phi_{b|q}$ is the difference in angle defined in (\ref{eq:selected beamforming_gain}). {It is verified in \cite{Ref_Jin08} that each  error follows $\beta(M-N,N)$ random variable and its cumulative distribution function (cdf) can be approximated for small $s$, with $\delta= \binom{M-1}{N-1}^{\frac{-1}{M-N}}$, according to}
\begin{align}
\label{eq:approximated_cdf}
\mathrm{F}_{\mathrm{S}}(s) \simeq \binom{M-1}{N-1}s^{M-N}  \mathbbm{1}_{[0,\delta]}(s) +\mathbbm{1}_{(\delta,1]}(s).
\end{align}

{
The \textit{local} CDI is obtained by selecting the codeword corresponding to the smallest quantization error from among ${Q}^{\mathrm{cl}}$ error terms $\sin^{2}\phi_{b|q},~q \in \{1,\cdots, {Q}^{\mathrm{cl}} \}
$.
%
The channel quantization error corresponding to the  \textit{local} CDI  can be studied by deriving the distribution of the  smallest quantization error
\begin{align*}
 \sin^2\phi_b & \doteq \sin^2\phi_{b|\hat{q}},
\end{align*}
{where the index of the  codeword in (\ref{eq:sel_local}) is rewritten by}
\begin{align}
\label{eq:sel_index}
\hat{q}&=\argmin_{q \in \{1,\cdots,{Q}^{\mathrm{cl}} \}} \sin^2\phi_{b|q}.
\end{align}
{Based on the largest order statistics, we derive the expectation of the smallest quantization error in the following proposition.}}
\begin{prop}
\label{pr:quantization_error}
{The expectation of the channel quantization error corresponding to the  \textit{local} CDI is {approximated by}}
\begin{align*}
\mathrm{E}[\sin^2\phi_b] \simeq {Q}^{\mathrm{cl}} \binom{M-1}{N-1}^{\frac{-1}{M-N}} \mathbf{B}\bigg({Q}^{\mathrm{cl}},\frac{M-N+1}{M-N}\bigg).
\end{align*}
\end{prop}
\begin{IEEEproof}
{Minimizing the quantization error is {the} same as maximizing the normalized beamforming gain}
\begin{align*}
\mathrm{C} \doteq \cos^{2}\phi_{b|q}&=\big| (\bd^q)^H {\bh}_{b|q}^{\mathrm{virt}} /\| {\bh}_{b|q}^{\mathrm{virt}} \|_2 \big|^2
\end{align*}
The cdf of the normalized beamforming gain is  given by
\begin{align*}
&\mathrm{F}_{\mathrm{C}}(c) 
= 1-\mathrm{F}_{\mathrm{S}}(1-c)
\\
&\simeq   1- \bigg(\binom{M-1}{N-1}(1-c)^{M-N}  \mathbbm{1}_{[0,\delta]}(1-c) +\mathbbm{1}_{(\delta,1]}(1-c) \bigg)
\\
&\stackrel{(a)} =  1- \bigg(\binom{M-1}{N-1}(1-c)^{M-N}  \mathbbm{1}_{[1-\delta,1]}(c) + \mathbbm{1}_{[0,1-\delta)}(c) \bigg)
\\
&\stackrel{(b)} = \bigg(1- \binom{M-1}{N-1}(1-c)^{M-N}\bigg) \mathbbm{1}_{[1-\delta,1]}(c),
\end{align*}
{where $(a)$ is derived because $\mathbbm{1}_{[0,\delta]}(1-c)=\mathbbm{1}_{[1-\delta,1]}(c)$, $\mathbbm{1}_{(\delta,1]}(1-c)=\mathbbm{1}_{[0,1-\delta)}(c)$, and $(b)$ is derived because $\mathrm{F}_{\mathrm{C}}(c)=0$ when $0 \leq c < 1-\delta$.}

\setcounter{equation}{27}
{The normalized beamforming gain corresponding to the selected codeword (\textit{local} CDI) is  written {as}
\begin{align*}
\mathrm{G} & \doteq \cos^2\phi_b= \max_{q \in \{1,\cdots,{Q}^{\mathrm{cl}} \}} \cos^{2}\phi_{b|q}.
\end{align*}
The probability that the largest beamforming gain is smaller than an arbitrary number $g$ is defined according to $\mathrm{Pr}\big(\cos^2\phi_b<g\big) = \prod_{q=1}^{{Q}^{\mathrm{cl}}} \mathrm{Pr}\big( \cos^2\phi_{b|q}<g\big)$.
Therefore, the cdf of $\mathrm{G}$ is  defined with the cdf of  $\mathrm{C}$ such as}
\begin{align*}
\mathrm{F}_{\mathrm{G}}(g)&\doteq \big(\mathrm{F}_{\mathrm{C}}(g)\big)^{{Q}^{\mathrm{cl}}}
\\
& \simeq \bigg(1- \binom{M-1}{N-1}(1-g)^{M-N}\bigg)^{{Q}^{\mathrm{cl}}} \mathbbm{1}_{[1-\delta,1]}(g).
\end{align*}
{Because $\mathrm{G}$ is a non-negative random variable, the expectation of $\mathrm{G}$ can be derived such that}
\begin{align*}
\nonumber
&{\mathrm{E}[\mathrm{G}] \simeq 1-\int_{0}^{1} \mathrm{F}_{\mathrm{G}}(g) dg}
\\
&=1-\sum_{q=0}^{{Q}^{\mathrm{cl}}}\binom{{{Q}^{\mathrm{cl}}}}{q}(-1)^{q}\binom{M-1}{N-1}^{q}\int_{1-\delta}^{1}{(1-g)}^{q(M-N)}dg
\\
\nonumber
&=1-\delta\sum_{q=0}^{{{Q}^{\mathrm{cl}}}}\frac{\binom{{{Q}^{\mathrm{cl}}}}{q}(-1)^{q}}{q(M-N)+1}
\\
\nonumber
&\stackrel{(a)}=1-\delta \sum_{q=0}^{{{Q}^{\mathrm{cl}}}}{\frac {(-{{Q}^{\mathrm{cl}}})_{q}} {q! \big(q+\frac{1}{M-N}\big)} \Big(\frac{1}{M-N}\Big)}
\\
\nonumber
&=1-\delta\frac{\Gamma {\big(\frac{1}{M-N}+1\big)}{{Q}^{\mathrm{cl}}}({{Q}^{\mathrm{cl}}}-1)!  }{\Gamma{\big(\frac{1}{M-N}+{{Q}^{\mathrm{cl}}}+1\big)}}
\\
\nonumber
&=1- {{Q}^{\mathrm{cl}}}\binom{M-1}{N-1}^{\frac{-1}{M-N}} \mathbf{B}\bigg({{Q}^{\mathrm{cl}}},\frac{M-N+1}{M-N}\bigg),
\end{align*}
where $(a)$ is derived based on \cite[6.6.8]{Ref_Han75}.

\begin{figure}
\centering
\subfigure{\includegraphics[width=0.425\textwidth]{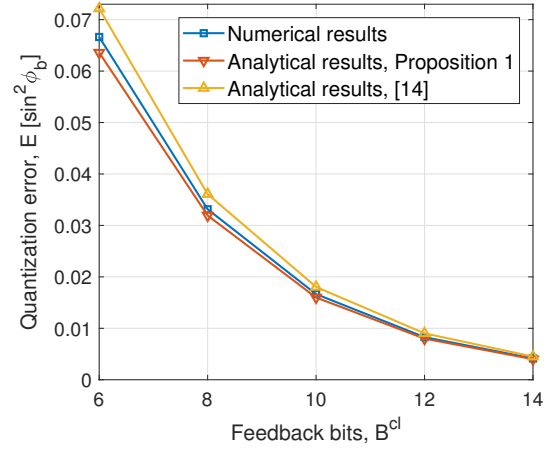}}
\caption{Quantization error for QBC $(M=4, N=2)$.}
\label{fig:verify}
\end{figure}

\begin{figure*}[!t]
\normalsize
\centering
\subfigure[{\textit{Local}  quan. error}]{\label{fig:quan_error_AU}\includegraphics[width=0.1515\textwidth]{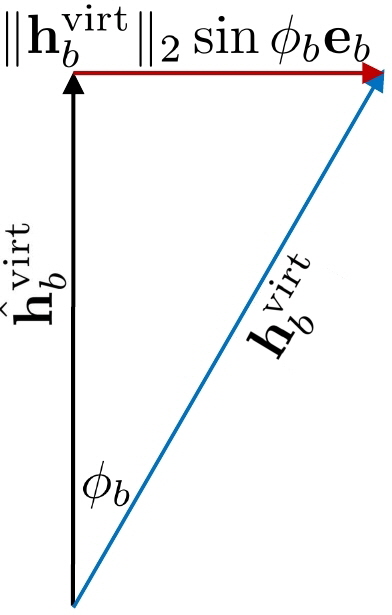}}
\hfil
\subfigure[{\textit{Global} quan. error}]{\label{fig:quan_error_MU}\includegraphics[width=0.215\textwidth]{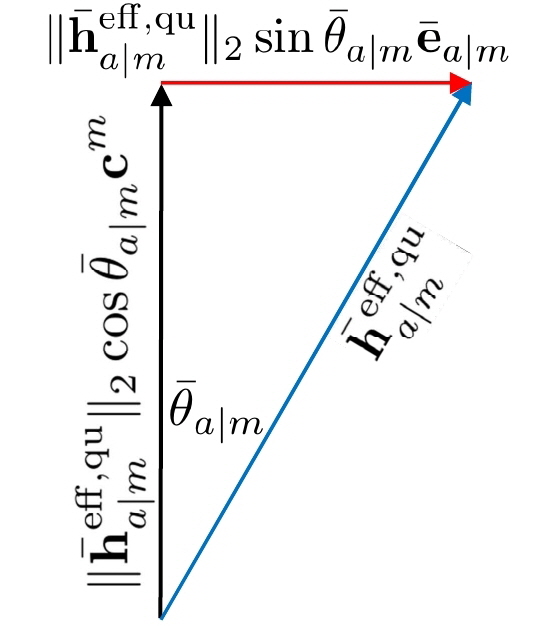}}
\hfil
\subfigure[{\textit{Local} and \textit{Global}  quan. errors}]{\label{fig:quan_error_CU}\includegraphics[width=0.215\textwidth]{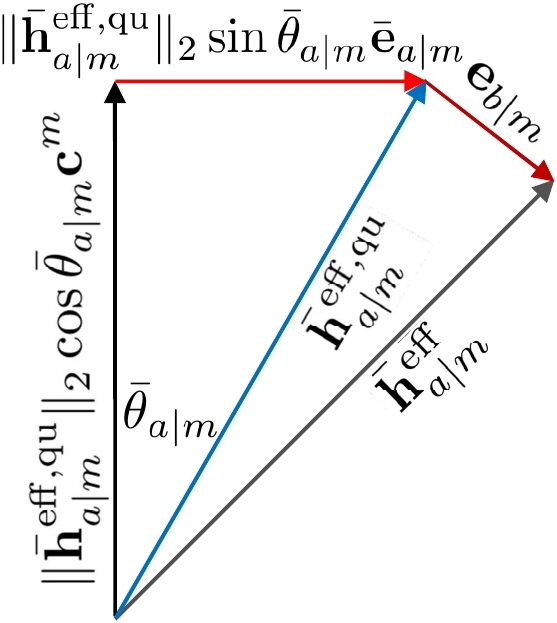}}
\caption{Channel quantization errors in cooperative feedback algorithm.}
\label{fig:error}
\hrulefill
\begin{align}
\setcounter{equation}{34}
\label{eq:rewritten}
y_{a|m}=\sqrt{\frac{\rho}{M}}\bigg(  \underbrace{ \big( \|\bar{\bh}_{a|m}^{\mathrm{eff,qu}}\|_2 \cos \bar{\theta}_{a|m}\bc^m+{\bee}_{b|m}\big)^H\bc^m }_{(I)}s_m  +     \sum_{\substack {\ell\ne m}}^M  \underbrace{   \| \bar{\bh}_{a|m}^{\mathrm{eff,qu}} \|_2  \sin\bar{\theta}_{a|m} \bar{\bee}_{a|m}^H\bc^\ell }_{(II)} s_{\ell}  +  \sum_{\substack {\ell \ne m}}^M    \underbrace{{\bee}_{b|m}^H\bc^\ell  }_{(III)}s_{\ell}  \bigg) + \underbrace{ \bar{n}_{a|m}}_{(IV)}.
\end{align}
\hrulefill
\end{figure*}

{The expectation of the quantization error corresponding to the \textit{local} CDI  is approximated such that
\begin{align*}
\mathrm{E} [ \sin^{2}\phi_b ]& \simeq 
{{Q}^{\mathrm{cl}}} \binom{M-1}{N-1}^{\frac{-1}{M-N}} \mathbf{B}\bigg({Q}^{\mathrm{cl}},\frac{M-N+1}{M-N}\bigg),
\end{align*}
because $\sin^{2}\phi_b=1-\cos^{2}\phi_b$. This completes the proof.}
\end{IEEEproof}

{Finally, the accuracy of the  quantization error  in Proposition \ref{pr:quantization_error} and the  formulation in \cite{Ref_Jin08}, i.e.,
$\big[{Q}^{\mathrm{cl}}\binom{M-1}{N-1}\big]^{\frac{-1}{M-N}}$, are evaluated by numerical results in Fig. \ref{fig:verify}. It is shown that our formulation in Proposition \ref{pr:quantization_error} shows better accuracy than the  formulation in \cite{Ref_Jin08}. Further, we point out that the quantization error is better fitted to the numerical results as  the feedback bits ${B}^{\mathrm{cl}}$ increase because the  cdf in (\ref{eq:approximated_cdf}) is valid for small $s$.}

\subsection{Received Signal of MU}
\label{sec:received_signal}

\setcounter{equation}{27}
{We take a closer look at the received  signal of MU in (\ref{eq:re_combine}). Assuming MU $a$ is scheduled to use the $m$-th beamformer  {$\bc^m$ (meaning that $\bar{u}^{m}=a$}), the received signal is written by
\begin{align}
\label{eq:re_write}
&\bar{y}_{a|m}=\sqrt{\rho}\bar{\bz}_{a|m}^H\bar{\bH}_a \bigg( \frac{[{\bc}^{1},\cdots,{\bc}^{M}]}{\sqrt{M}} \bigg) \bs + \bar{\bz}_{a|m}^H\bar{\bn}_{a}
\\
\nonumber
&=\sqrt{\frac{\rho}{M}} \bigg( (\bar{\bh}_{a|m}^{\mathrm{eff}})^H  \bc^m  s_m+  \sum_{\substack {\ell=1,\ell \ne m}}^M   (\bar{\bh}_{a|m}^{\mathrm{eff}})^H \bc^\ell  s_{\ell} \bigg) +\bar{n}_{a|m},
\end{align}
where $\bar{\bh}_{a|m}^{\mathrm{eff}} \doteq   \bar{\bH}_a^H \bar{\bz}_{a|m}$ is the  \textit{global} effective channel vector, and $\bar{n}_{a|m} \doteq \bar{\bz}_{a|m}^H\bar{\bn}_{a} \sim \mathcal{CN}(0,1)$ is the combined noise.
In order to examine \textit{beamforming gain} and \textit{interuser interference}, we must investigate the cross-correlations between  the  \textit{global} effective channel vector $\bar{\bh}_{a|m}^{\mathrm{eff}}$ and   codewords in $\mathcal{C}$.}

\setcounter{equation}{28}
{The  \textit{global} effective channel vector is computed through two antenna combining processes (\textit{locally} in AU and \textit{globally} in MU) and each antenna combining process causes an individual quantization error.
Before investigating both quantization errors jointly, we discuss each quantization error separately.
First, we  discuss the channel quantization error caused in the process of  virtual vector quantization using \textit{local} codebook $\mathcal{D}$. 
Note that the \textit{local} quantization error is presented in Section \ref{sec:local_comb}.
As illustrated in Fig. \ref{fig:quan_error_AU}, the  virtual channel vector ${\bh}_{b}^{\mathrm{virt}}$ in (\ref{eq:signal_AU}) is divided into the quantized virtual channel vector $\hat{\bh}_{b}^{\mathrm{virt}}$ in (\ref{eq:quan_virtual}) and the \textit{local} error vector  ${\bee}_b$ according to
\begin{align}
\label{eq:error_AU}
{\bh}_{b}^{\mathrm{virt}}  &= ({\| {\bh}_{b}^{\mathrm{virt}} \|_2} \cos\phi_{b})  {\bv}_b +  ({\| {\bh}_{b}^{\mathrm{virt}} \|_2} \sin\phi_{b} ) {\bee}_b
\\
\nonumber
&=  \hat{\bh}_{b}^{\mathrm{virt}}  + ({\| {\bh}_{b}^{\mathrm{virt}} \|_2}\sin\phi_{b} ) {\bee}_b,
\end{align}
%
where $\sin^2\phi_{b} = 1- \big|{\bv}_b^H {\bh}_{b}^{\mathrm{virt}}/\|{\bh}_{b}^{\mathrm{virt}} \|_2 \big|^2$
quantifies  the \textit{local} quantization error.}
{{Second, we discuss the channel quantization error caused in the process of  effective vector quantization using the \textit{global} codebook $\mathcal{C}$.}  
Note that the \textit{global} quantization error is presented in Section \ref{sec:global_comb}.
As depicted in Fig. \ref{fig:quan_error_MU},  the effective channel vector $\bar{\bh}_{a|m}^{\mathrm{eff,qu}}$  in (\ref{eq:effective_vector}) is divided into the  target codeword  $\bc^m$ and the \textit{global} error vector  $\bar{\bee}_{a|m}$  such that
\begin{align}
\label{eq:sel_error_CU}
\bar{\bh}_{a|m}^{\mathrm{eff,qu}}=  (\|  \bar{\bh}_{a|m}^{\mathrm{eff,qu}}\|_2  \cos\bar{\theta}_{a|m})  \bc^m + (\|  \bar{\bh}_{a|m}^{\mathrm{eff,qu}}\|_2 \sin\bar{\theta}_{a|m} ) \bar{\bee}_{a|m},
\end{align}
where $\sin^2\bar{\theta}_{a|m} =1- \big| (\bc^m)^H \bar{\bh}_{a|m}^{\mathrm{eff,qu}}/\| \bar{\bh}_{a|m}^{\mathrm{eff,qu}} \|_2 \big|^2$ quantifies the \textit{global} quantization error.}

\begin{figure*}[!t]
\setcounter{equation}{35}
\begin{align}
\nonumber
{\textrm{SINR}}_{a|m} & \geq \frac{    \frac{\rho}{M}\| \bar{\bh}_{a|m}^{\mathrm{eff,qu}}\|_2^2 \cos^2 \bar{\theta}_{a|m}  }{ \mathrm{E}[|\bar{n}_{a|m}|^2] + \frac{\rho}{M}\| \bar{\bh}_{a|m}^{\mathrm{eff,qu}} \|_2^2  \sin^2\bar{\theta}_{a|m}   \sum_{{\ell=1, \ell \ne m}}^M   \underbrace{ | \bar{\bee}_{a|m}^H\bc^\ell|^2}_{(i)} + \frac{\rho}{M} |[\bar{\bz}_{a|m}]_{N+1}|^2 \| {\bh}_{b}^{\mathrm{virt}} \|_2^2 \sin^2\phi_{b}  \sum_{\substack {\ell=1}}^M  \underbrace{ |{\bee}_b^H\bc^{\ell}|^2 }_{(ii)} }
\\
\nonumber
& =\frac{    \frac{\rho}{M}\| \bar{\bh}_{a|m}^{\mathrm{eff,qu}}\|_2^2 \cos^2 \bar{\theta}_{a|m}  }{  1 + \frac{\rho}{M}\| \bar{\bh}_{a|m}^{\mathrm{eff,qu}} \|_2^2  \sin^2\bar{\theta}_{a|m} + \frac{\rho}{M} |[\bar{\bz}_{a|m}]_{N+1}|^2 \| {\bh}_{b}^{\mathrm{virt}} \|_2^2 \sin^2\phi_{b}  \sum_{\substack {\ell=1}}^M  |{\bee}_b^H\bc^{\ell}|^2 }
\\
\label{eq:SINR}
& =\frac{    \frac{\rho}{M}\| \bar{\bh}_{a|m}^{\mathrm{eff,qu}}\|_2^2 \cos^2 \bar{\theta}_{a|m}  }{  1 + \frac{\rho}{M} \underbrace{ \| \bar{\bh}_{a|m}^{\mathrm{eff,qu}} \|_2^2  \sin^2\bar{\theta}_{a|m}}_{(II)}  + \frac{\rho}{M}  \underbrace{|[\bar{\bz}_{a|m}]_{N+1}|^2 \| {\bh}_{b}^{\mathrm{virt}} \|_2^2 \sin^2\phi_{b}}_{(III)}   }.
\end{align}
\hrulefill
\end{figure*}



\setcounter{equation}{30}
{We now consider both quantization errors   jointly.  When MU conducts post-signal processing, the \textit{global} combiner $\bar{\bz}_{a|m}$ is used to combine spatial dimensions of  the \textit{global} channel matrix $\bar{\bH}_a$ in (\ref{eq:virtual_matrix_trans}) for downlink transmissions. {We call $\bar{\bH}_a$ the downlink channel matrix.}  It should be noted that the downlink channel matrix $\bar{\bH}_a$ includes   the \textit{unquantized} virtual vector $\bh_b^{\mathrm{virt}}$ in (\ref{eq:unquan_local}). However, the \textit{global} combiner is computed   using another \textit{global} channel matrix   $\bar{\bH}^{\mathrm{qu}}_a$ in (\ref{eq:virtual_matrix}) including the \textit{quantized} virtual  vector $\hat{\bh}_b^{\mathrm{virt}}$ in (\ref{eq:quan_virtual}).
The correspondence between the  downlink channel matrix   and the \textit{global} channel matrix  must be investigated because the combined quantization error occurs due to the difference between  $\bar{\bH}_a$  and $\bar{\bH}^{\mathrm{qu}}_a$.}

{First, we rewrite the  downlink channel matrix in (\ref{eq:virtual_matrix_trans}) by plugging the virtual channel vector in (\ref{eq:error_AU})   according to}
\begin{align}
\nonumber
{\bar{\bH}_a} &= \bigg[\begin{array}{c}
                \bH_a \\
                ({\bh}_{b}^{\mathrm{virt}})^H
\end{array}\bigg]
\\
\nonumber
&
= \bigg[\begin{array}{c}
                \bH_a \\
                (\hat{\bh}_{b}^{\mathrm{virt}})^H
\end{array}\bigg]
+
\bigg[\begin{array}{c}
                \mathbf{0}_{N,M} \\
                ({\| {\bh}_{b}^{\mathrm{virt}} \|_2}\sin\phi_{b})  {\bee}_b^H
\end{array}\bigg]
\\
\label{eq:relation}
&=\bar{\bH}^{\mathrm{qu}}_a+ \big[\mathbf{0}_{M,N},~ ({\| {\bh}_{b}^{\mathrm{virt}} \|_2}\sin\phi_{b} ) {\bee}_b \big]^H.
\end{align}
{As depicted in Fig. \ref{fig:quan_error_CU}, the  effective channel vector in (\ref{eq:re_write}),  corresponding to downlink transmissions,   is written by\footnote{{We call $\bar{\bh}_{a|m}^{\mathrm{eff}}$  the downlink channel vector.}}
\begin{align}
\nonumber
\bar{\bh}_{a|m}^{\mathrm{eff}} &={\bar{\bH}_a}^H \bar{\bz}_{a|m}
\\
\nonumber
&= (\bar{\bH}^{\mathrm{qu}}_a)^H  \bar{\bz}_{a|m} + \big[\mathbf{0}_{M,N},~ ({\| {\bh}_{b}^{\mathrm{virt}} \|_2}\sin\phi_{b} ) {\bee}_b \big] \bar{\bz}_{a|m}
\\
\label{eq:induced_error_AU}
&= \bar{\bh}_{a|m}^{\mathrm{eff,qu}}+{\bee}_{b|m},
\end{align}
where the effective channel vector  $\bar{\bh}_{a|m}^{\mathrm{eff,qu}}=(\bar{\bH}^{\mathrm{qu}}_a)^H  \bar{\bz}_{a|m}$ is defined in (\ref{eq:effective_vector}) and the error vector is given by}
\begin{align}
\label{eq:induced_error}
{\bee}_{b|m} \doteq ([\bar{\bz}_{a|m}]_{N+1} \| {\bh}_{b}^{\mathrm{virt}} \|_2 \sin\phi_{b}) {\bee}_b \in \mathbb{C}^M.
\end{align}

{Second, we rewrite the downlink channel vector   $\bar{\bh}_{a|m}^{\mathrm{eff}}$   by plugging the effective channel vector $\bar{\bh}_{a|m}^{\mathrm{eff,qu}} $ in (\ref{eq:sel_error_CU}) such that
\begin{align}
\nonumber
\bar{\bh}_{a|m}^{\mathrm{eff}} &= \underbrace{    (\| \bar{\bh}_{a|m}^{\mathrm{eff,qu}} \|_2\cos\bar{\theta}_{a|m})\bc^m}_{(a)}
\\
\label{eq:error_total}
&\,\,\,\,\,\, + \underbrace{(\| \bar{\bh}_{a|m}^{\mathrm{eff,qu}} \|_2 \sin\bar{\theta}_{a|m} ) \bar{\bee}_{a|m}}_{(b)}  + \underbrace{{\bee}_{b|m}}_{(c)},
\end{align}
where $\| \bar{\bh}_{a|m}^{\mathrm{eff,qu}} \|_2^2\cos^2\bar{\theta}_{a|m}$ in $(a)$ denotes  the beamforming gain for data transmissions, $(b)$ denotes the \textit{global} quantization error, and $(c)$ denotes the \textit{local} quantization error. The  relationship between the codeword $\bc^m$ and the \textit{global} effective  vector $\bar{\bh}_{a|m}^{\mathrm{eff}}$ is summarized in Fig. \ref{fig:quan_error_CU}.}

{Finally, we rewrite the received signal of MU  by plugging the downlink  vector $\bar{\bh}_{a|m}^{\mathrm{eff}}$ in (\ref{eq:error_total}) into the input-output expression in  (\ref{eq:re_write}). As presented in (\ref{eq:rewritten}), the desired signal  $(I)$,  \textit{global} (interuser) interferences $(II)$, \textit{local} (interuser) interferences $(III)$, and noise $(IV)$ {are clearly distinguished.}}

\subsection{Analysis of signal-to-interference-plus-noise ratio}
\label{sec:cqi_each_cu}


{We  analyze the SINR for  all MUs  in the network. 
Similar to \cite{Ref_Yoo07}, the lower bound of SINR is defined in  (\ref{eq:SINR})  because $\sum_{\substack {\ell=1,\ell \ne m}}^M   |{\bee}_b^H\bc^{\ell}|^2=\sum_{\substack {\ell=1}}^M   |{\bee}_b^H\bc^{\ell}|^2$  based on the assumption that the \textit{local} quantization error ${\bee}_{b|m}$ is orthogonal to the target codeword $\bc^m$.
We noted that  the error vector ${\bee}_{b|m}$  is replaced by (\ref{eq:induced_error}) in order to show  all the   \textit{local} interference terms  clearly.
To verify the second equality, we should recall 
%
that the unit-norm error vector  $\bar{\bee}_{a|m}$  is orthogonal to the target codeword $\bc^m$.
Since we consider a single unitary matrix for constructing the \textit{global} codebook,\footnote{{If  \textit{global} codebook consists of more than $M$ codewords,  $| \bar{\bee}_{a|m}^H \bc^{\ell} |^2$  can be modeled by $\beta(1,M-2)$ because $\bar{\bee}_{a|m}$ and $\bc^{\ell}$ independent and isotropically distributed on the $M-1$ dimensional hyperplane orthogonal to $\bc^m$ \cite{Ref_Jin06}.}}  
the error vector   is on the hyperplane, which is  made up of $M-1$ orthonormal codewords that are orthogonal to  $\bc^m$.
The error vector is thus written as a linear combination of the orthonormal codewords  as
$\bar{\bee}_{a|m} = \sum_{{\ell=1, \ell \ne m}}^M ((\bc^\ell)^H\bar{\bee}_{a|m})\bc^\ell$.
The sum of terms in  $(i)$ is computed according to
$
\| \bar{\bee}_{a|m} \|_2^2=\sum_{{\ell=1, \ell \ne m}}^M  |\bar{\bee}_{a|m}^H\bc^\ell|^2 =1,
$
because $\bc^{\ell}$ and $\bar{\bee}_{a|m}$ are unit-norm vectors.
The third equality follows because the sum of  terms in $(ii)$ is computed as
\begin{align*}
\sum_{\ell=1}^M  |{\bee}_b^H\bc^{\ell}|^2={\bee}_b^H(\bC \bC^H) {\bee}_b =\|{\bee}_b\|_2^2=1,
\end{align*}
because $\bC \doteq [\bc^1,\cdots, \bc^M] \in \mathbb{C}^{M \times M}$ is the unitary matrix.

\begin{figure}
\centering
\subfigure[Distributions of \textit{local} and \textit{global} quan. errors]{\label{fig:cdf_variance}\includegraphics[width=0.445\textwidth]{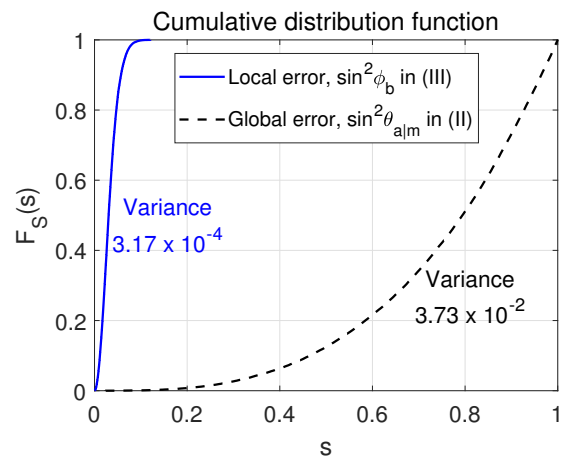}}
\subfigure[Distributions of \textit{local} and \textit{global} interferences]{\label{fig:int_variance}\includegraphics[width=0.445\textwidth]{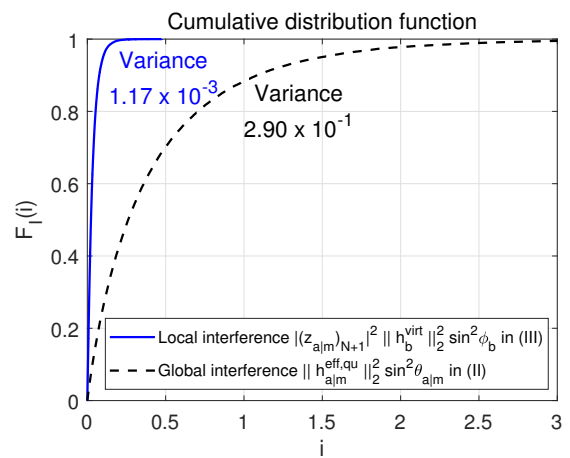}}
\caption{CDF of \textit{local} and \textit{global} variables ($M=4,~N=2,~{B}^{\mathrm{cl}}=8$).}
\end{figure}

\begin{figure*}[!t]
\setcounter{equation}{36}
\begin{align}
\label{eq:fin_SINR}
 {\textrm{SINR}}_{a|m} & \simeq    \frac{    \frac{\rho}{M} \| \bar{\bh}_{a|m}^{\mathrm{eff,qu}}\|_2^2 \cos^2 \bar{\theta}_{a|m}  }{   1  +  \frac{\rho}{M}\| \bar{\bh}_{a|m}^{\mathrm{eff,qu}} \|_2^2  \sin^2\bar{\theta}_{a|m} +\frac{\rho}{M}   \mathrm{E} \big[ |[\bar{\bz}_{a|m}]_{N+1}|^2 \| {\bh}_{b}^{\mathrm{virt}} \|_2^2 \sin^2\phi_{b} \big]   }
\simeq   \frac{     \frac{\rho}{M \alpha}\| \bar{\bh}_{a|m}^{\mathrm{eff,qu}}\|_2^2 \cos^2 \bar{\theta}_{a|m}  }{ 1   +    \frac{\rho}{M \alpha}\| \bar{\bh}_{a|m}^{\mathrm{eff,qu}}\|_2^2  \sin^2 \bar{\theta}_{a|m}    } \doteq \bar{\gamma}_{a|m}.
\end{align}
\hrulefill
\setcounter{equation}{38}
\begin{align}
\label{eq:selected_CQI}
\bar{\gamma}_{|m}\simeq \frac{\rho \varrho^{2}}{ M \alpha } \bigg(\log\bigg({\frac{\binom{M-1}{N}\bar{Q}^{m}}{\big( \frac{\rho \varrho^{2} }{ M \alpha }\big)^{(M-(N+1))}}}\bigg)-(M-(N+1))\log{\bigg(\log\bigg({\frac{\binom{M-1}{N}\bar{Q}^{m}}{\big(\frac{\rho \varrho^{2}}{ M \alpha }\big)^{(M-(N+1))}}}\bigg) + \Big(\frac{\rho \varrho^{2}}{ M \alpha  } \Big)^{-1}\bigg) \bigg)}.
\end{align}
\hrulefill
\begin{align}
\setcounter{equation}{40}
\label{eq:selected_CQI_remark}
\gamma_{|m} \simeq  &\frac{\rho}{M}\bigg(\log\bigg({\frac{\binom{M-1}{N-1}Q^{m}}{ (\rho/M)^{(M-N)}}}\bigg)-(M-N)\log{\bigg(\log\bigg({\frac{\binom{M-1}{N-1}Q^{m}}{(\rho/M)^{(M-N)}}}\bigg) +\Big(\frac{\rho}{M}\Big)^{-1} \bigg)\bigg)}.
\end{align}
\hrulefill
\end{figure*}

Unfortunately, it  is   not easy to derive the joint distribution between the \textit{local} and \textit{global}  error terms in (\ref{eq:SINR}).
For easy of analysis, we  point out the fact that the  size of the \textit{local} codebook is much bigger than that of the \textit{global} codebook such as ${B}^{\mathrm{cl}} \gg B$.
Moreover,  $\sin^2\phi_{b}$ is the minimum error selected among $Q^{\mathrm{cl}}$ \textit{local} error terms, while  $\cos^2 \bar{\theta}_{a|m}$ is  an general \textit{global} error term.
For these reasons, the variance and magnitude of the  \textit{local} quantization error  are smaller than those of the \textit{global} quantization error, as shown in Fig. \ref{fig:cdf_variance}.
It  is also verified in Fig. \ref{fig:int_variance} that the variance and magnitude of the \textit{local} interference are small compared to those of the \textit{global} interference.
%
{From these observations, we infer that the \textit{local} interferences  would have  minimal effect on the  SINR.}

\begin{figure}
\centering
\subfigure{\includegraphics[width=0.445\textwidth]{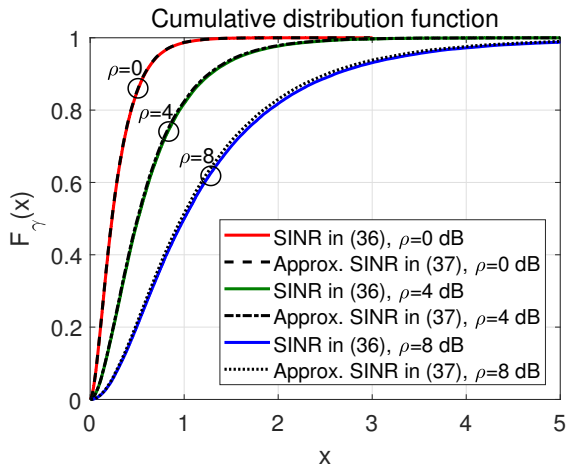}}
\caption{CDF of SINR and approximated SINR ($M=4,~N=2,~{B}^{\mathrm{cl}}=8$).}
\label{fig:cdf_sinr}
\end{figure}

{In this paper, we approximate the SINR by computing the expectation of \textit{local} interference using similar logic that the absolute square of the noise term is commonly replaced by its expectation value when defining the SINR. We  noted that the accuracy of this approximation is evaluated in the following paragraph.}
{The expectation of  \textit{local} interference  is
\begin{align*}
\mathrm{E} \big[|[\bar{\bz}_{a|m}]_{N+1} |^2 \| {\bh}_{b}^{\mathrm{virt}} \|_2^2  \sin^2 \phi_{b}\big]& \stackrel{(a)}=\frac{   \mathrm{E} [ \| {\bh}_{b}^{\mathrm{virt}} \|_2^2 ] \mathrm{E}  [ \sin^2 \phi_{b} ]    }{N+1}
\\
&\stackrel{(b)}=\frac{(M-N+1) \mathrm{E}  [ \sin^2 \phi_{b} ]}{N+1}
\\
&\stackrel{(c)}\simeq\frac{(M-N+1) \omega  }{N+1}\doteq \nu,
\end{align*}
because  $|[\bar{\bz}_{a|m}]_{N+1} |^2$, $\| {\bh}_{b}^{\mathrm{virt}} \|_2^2 $, and  $ \sin^2 \phi_{b} $ are    independent. Note that $(a)$ is derived because the  absolute square of each entry of the  \textit{global} combiner $\bar{\bz}_{a|m} \in \mathbb{C}^{N+1}$ is expected to be one over the number of entries, $(b)$ is derived because the quantity  $\| {\bh}_{b}^{\mathrm{virt}} \|^{2}_2$ of effective channel   follows Chi-squared distribution  $\chi^2_{2(M-N+1)}$ and its expectation is $M-N+1$  \cite[Lemma 3]{Ref_Jin08},  and $(c)$ is derived in Proposition \ref{pr:quantization_error}, where the expectation of the quantization error is $\omega ={Q}^{\mathrm{cl}}{\binom{M-1}{N-1}}^{\frac{-1}{M-N}}\mathbf{B}\big({Q}^{\mathrm{cl}},\frac{M-N+1}{M-N}\big)$.  The  SINR is  approximated  in (\ref{eq:fin_SINR}), where $\alpha \doteq 1+   \rho\nu/M$ denotes the  noise-plus-\textit{local} interference for a given SNR $\rho$.}

{We  evaluate the accuracy of our approximation by comparing the cdf of the approximated SINR in (\ref{eq:fin_SINR}) with that of the SINR in (\ref{eq:SINR}). As shown in Fig. \ref{fig:cdf_sinr}, the SINR is well approximated by (\ref{eq:fin_SINR}) in three different SNR scenarios. We thus conclude that  the operation of replacing the \textit{local} interference  by its expectation value would lead a minimal effect on the SINR. For easy of analysis, we use the approximated SINR for the rest of sections.}

{Further, the distribution of the quantity $\|\bar{\bh}_{a|m}^{\mathrm{eff,qu}}\|_2^2$ in (\ref{eq:fin_SINR}) is derived in the following proposition.
\begin{prop}
\setcounter{equation}{37}
\label{pr:effective_vector}
The squared norm of \textit{global} effective vector
\begin{align*}
\mathrm{U} \doteq  \| \bar{\bh}_{a|m}^{\mathrm{eff,qu}}\|_2^2=\| (\bar{\bH}^{\mathrm{qu}}_a)^H \bar{\bz}_{a|m}\|_2^2
\end{align*}
follows  Chi-squared distribution $\chi^2_{2(M-N)}$ and its pdf is
$f_{\mathrm{U}}(u) \simeq   \frac{u^{M-N-1} e^{-u/\varrho^2}}{ \varrho^{2(M-N)}\Gamma{(M-N)}}$
with the variance $\varrho^{-2} \doteq \frac{1}{N+1}\big[N+\frac{M}{(1-\omega)(M-N+1)}\big]$ and $\omega  \doteq {Q}^{\mathrm{cl}}{\binom{M-1}{N-1}}^{\frac{-1}{M-N}}\mathbf{B}\big({Q}^{\mathrm{cl}},\frac{M-N+1}{M-N}\big)$.
\end{prop}
\begin{IEEEproof}
For the proof, see Appendix \ref{sec:Appendix_A}.
\end{IEEEproof}

Finally, the cdf of  approximated SINR $\mathrm{X} \doteq \bar{\gamma}_{a|m}$ in (\ref{eq:fin_SINR}) is derived based on  the pdf of $\| \bar{\bh}_{a|m}^{\mathrm{eff,qu}}\|_2^2$ according to}
\begin{align}
\label{eq:27}
\mathrm{F}_{\mathrm{X}}(x) \simeq 1-\frac{\binom{M-1}{N}e^{-\frac{M \alpha  }{\rho \varrho^{2}}x }}{(x+1)^{M-N-1}}.
\end{align}
\begin{IEEEproof}
{For the proof, see \cite{Ref_Yoo07} and \cite[Lemma 3]{Ref_Son12}.}
\end{IEEEproof}

\subsection{Cooperation Mode Switching Algorithm}
\label{sec:adap_cooper}
{{We develop {a} cooperation mode switching algorithm  based on the expectation of a sum-rate.}
\setcounter{equation}{39}
To estimate sum-rate throughput, we derive   the cdf of the SINR for   the scheduled MUs in $\bar{\mathcal{U}}$.
{By using a similar logic from \cite[Theorem 1]{Ref_Son12}, the  SINR for the $m$-th selected MU is estimated as in (\ref{eq:selected_CQI}) with the cdf of SINR in (\ref{eq:27}).}
We now take a closer look at  $\bar{Q}^{m}$ meaning the  number of CQI candidates for the $m$-th user selection process.
Since each user generates $M$ CQIs, the total number of CQI candidates  is given by $KM$.
Because the user having the largest CQI  is selected {from the} remaining  CQI candidates, the scheduled user and the codeword is excluded for the following user selection process.
The number of CQI candidates in the $m$-th user selection process is defined by}
\begin{align*}
\bar{Q}^{m}\doteq   \big(K-2(m-1)\big)\big(M-(m-1)\big).
\end{align*}

{Finally, the sum-rate of the multiuser MIMO system relying upon the proposed cooperative feedback is estimated as
\begin{align}
\label{eq:29}
\mathrm{R}_{\textrm{prop}} \simeq \sum_{m=1}^{M}\log_{2}{\big(1+\bar{\gamma}_{|m}\big)}
\end{align}
with the estimated SINR for the scheduled MUs  in (\ref{eq:selected_CQI}).}

\begin{remark}
\label{rm:no_cooperative feedback_sum}
{When the cooperative feedback is not activated, the sum-rate of the multiuser MIMO system  is estimated in \cite{Ref_Son12} according to
$\mathrm{R}_{\textrm{conv}} \simeq \sum_{m=1}^{M}\log_{2}{(1+\gamma_{|m})}$.
The  SINR for  the scheduled user is defined in (\ref{eq:selected_CQI_remark}) and   the number of CQI candidates is given by}
$Q^{m}\doteq \big(K-(m-1)\big)\big(M-(m-1)\big)$.
\end{remark}

 \setcounter{equation}{41}
{In the proposed cooperation mode switching algorithm,  multiuser MIMO systems {activate} the  cooperative feedback mode when the differential sum-rate is positive such that}
\begin{align}
\label{eq:cross_al01}
\bigtriangleup \mathrm{R} & \doteq \mathrm{R}_{\textrm{prop}}-\mathrm{R}_{\textrm{conv}}>0.
\end{align}



\section{Numerical Results}
\label{sec:13}
In this section, we evaluate the  performance of  the  cooperative feedback algorithm based on the sum-rate defined as
\begin{align*}
\mathrm{R}_{\textrm{num}}\doteq \sum_{m=1}^M \log_2(1+\check{\gamma}_{|m}),
\end{align*}
where the SINR for the $m$-th scheduled user is computed {numerically} according to 
\begin{align*}
\check{\gamma}_{|m} \doteq \frac{  | (\bar{\bh}_{\bar{u}^m}^{\mathrm{eff}})^H   \bc^m |^2 }{\frac{M}{\rho}+ \sum_{\substack {\ell=1,\ell \ne m}}^M  | (\bar{\bh}_{ \bar{u}^m}^{\mathrm{eff}})^H  \bc^{\ell}   |^2}.
\end{align*}
{Note that $\bar{u}^m$ denotes the scheduled MU exploiting the  $m$-th codeword $\bc^{m}$. The sum-rate performance is evaluated numerically from Monte-carlo simulations with $10,000$ independent  channels (solid blue lines in Fig. \ref{fig:3}). Moreover, the sum-rate performance is verified  analytically with the  formulation derived in Section \ref{sec:adap_cooper} (dotted black lines in Fig. \ref{fig:3}).}

\begin{figure}
\centering
\subfigure[$M=4,~N=3,~{B}^{\mathrm{cl}}=4$]{\label{fig:num_b_N3}\includegraphics[width=0.477\textwidth]{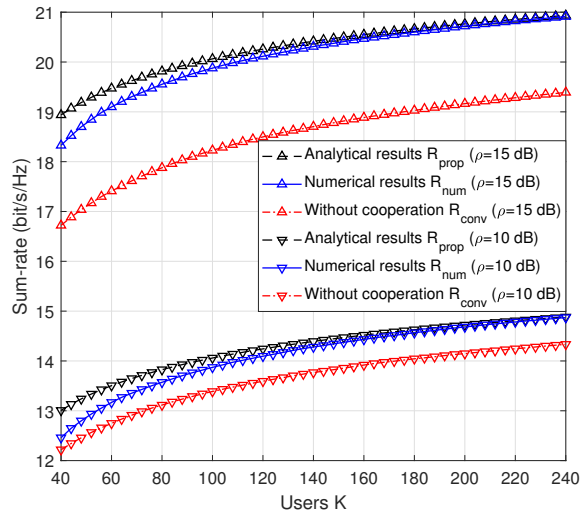}}
\subfigure[$M=4,~N=3,~{B}^{\mathrm{cl}}=4$]{\label{fig:num_anal_{k}ser_N3}\includegraphics[width=0.477\textwidth]{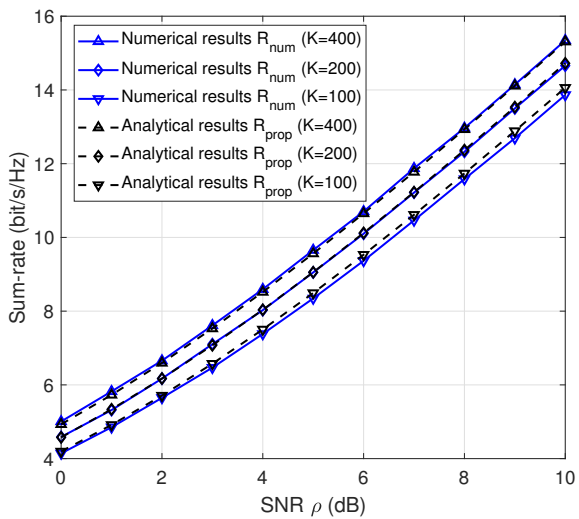}}
\subfigure[$M=4,~N=2,~{B}^{\mathrm{cl}}=8$]{\label{fig:num_anal_{k}ser_N2}\includegraphics[width=0.477\textwidth]{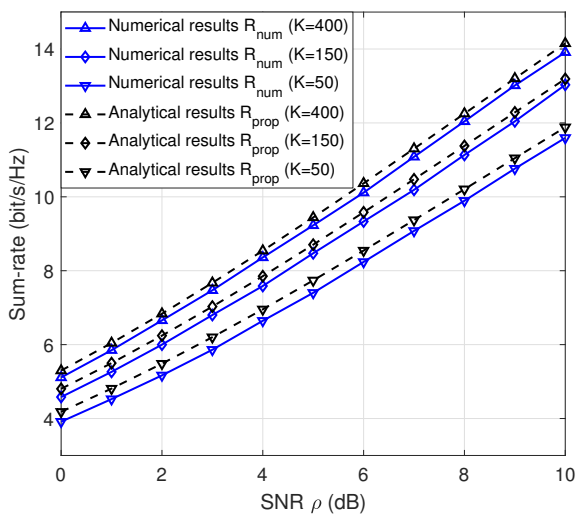}}
\caption{Comparison between  numerical and analytical results.}
\label{fig:3}
\end{figure}

{We first investigate the accuracy of the sum-rate formulation  $\mathrm{R}_{\textrm{prop}}$ derived in  (\ref{eq:29}). In Fig. \ref{fig:num_b_N3},  we  compare the numerical results and the sum-rate formulation  against the  number of users $K$.  In Figs. \ref{fig:num_anal_{k}ser_N3} and \ref{fig:num_anal_{k}ser_N2}, the numerical results and the sum-rate formulation  are compared {for a variety of user numbers} (between $50$ and  $400$) against  SNR $\rho$. The accuracy of the sum-rate formulation is verified by assuming the cooperation mode is activated. It should be noted that the sum-rate formulation is derived in Section \ref{sec:adap_cooper} based on the largest order statistics \cite{Ref_Dav80}. According to the extreme value theory, the differences between the numerical results and the sum-rate in  (\ref{eq:29})  decrease as the number of users $K$ increases. {In Figs. \ref{fig:3}, it is shown that the differences between the numerical results  and the sum-rate formulation are negligible, especially when there are {a large number of users on the MTC network.}} Moreover,  the error term $\nu $ in (\ref{eq:fin_SINR}) is matched to the numerical results when the overhead for the cooperation link ${B}^{\mathrm{cl}}$ is large. It is expected that the  sum-rate formulation {will} be  better fitted to the numerical results  as the size of the \textit{local} codebook ${B}^{\mathrm{cl}}$ increases.}

\begin{figure}
\centering
\subfigure[$M=4,~N=3,~{B}^{\mathrm{cl}}=6$]{\label{fig:sum_ava_N3}\includegraphics[width=0.454\textwidth]{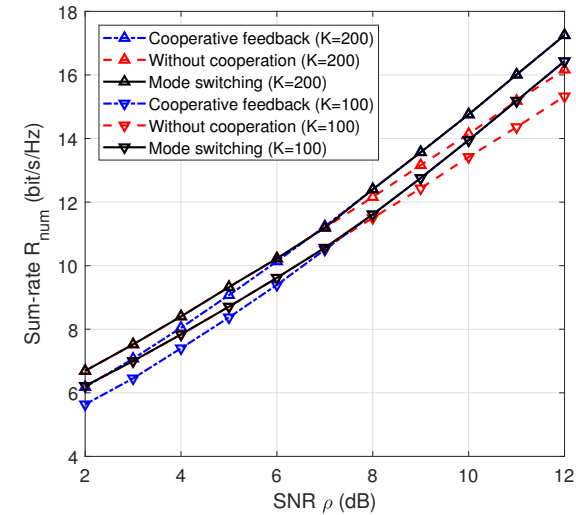}}
\subfigure[$M=4,~N=2,~{B}^{\mathrm{cl}}=4$]{\label{fig:sum_ava_N2}\includegraphics[width=0.454\textwidth]{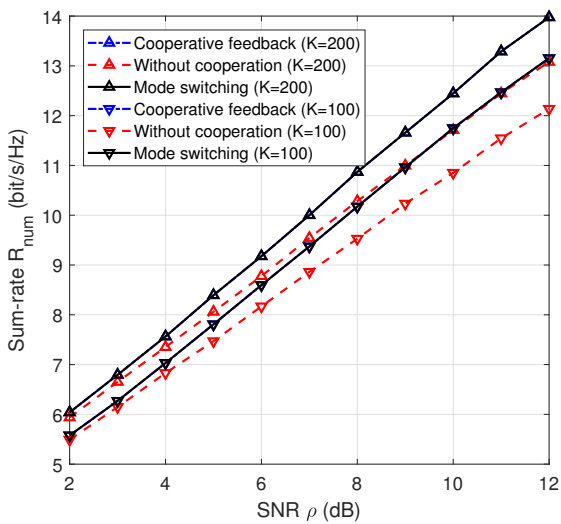}}
\subfigure[Mode switching  points in Fig. \ref{fig:sum_ava_N3}]{\label{fig:sum_ava_N3_closer}\includegraphics[width=0.454\textwidth]{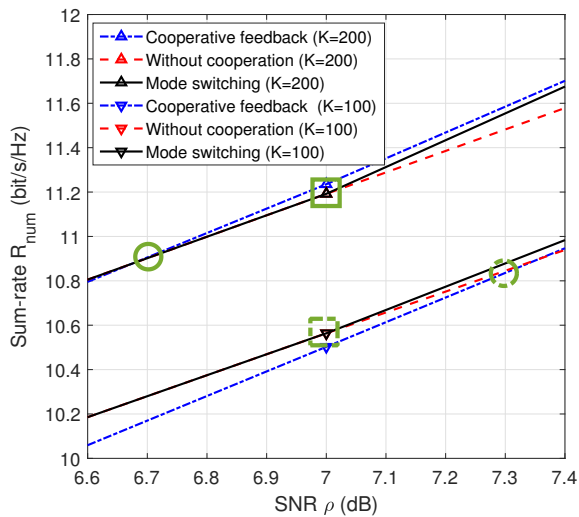}}
\caption{Sum-rate performance of adaptive cooperation  algorithm.}
\end{figure}

In Figs. \ref{fig:sum_ava_N3}  and \ref{fig:sum_ava_N2}, we  evaluate the sum-rate performances of the cooperation mode switching algorithm. {The proposed algorithm shows better sum-rate estimation performance when the number of users is {sufficiently large}   because the  cooperation mode switching algorithm is developed based on the extreme value theory \cite{Ref_Dav80}.} In Fig. \ref{fig:sum_ava_N3_closer}, we take a closer look at the cross point between the cooperative feedback activation mode and the cooperative feedback deactivation mode ($M=4,~N=3,~{B}^{\mathrm{cl}}=6$).
{The estimated mode switching point (square) and the mode switching point in the numerical results (circle) are both within the range of {a $0.3$-dB window.} This result means that the proposed adaptive cooperation algorithm finds mode switching points well based on the given information of SNR $\rho$ and system parameters, i.e., $K$, ${B}^{\mathrm{cl}}$, $M$, and $N$.}
In Fig. \ref{fig:sum_ava_N2},  the mode switching algorithm always triggers cooperation mode. For this reason, {the} cooperative feedback algorithm (blue line) and  mode switching algorithm (black line) produce the same numerical results and the numerical results {overlap}. Numerical simulations verify  that the cooperative feedback algorithms outperform conventional multiuser MIMO systems {that} do not exploit cooperative feedback mode.


\section{Conclusion}
\label{sec:14}
{In this paper, we developed limited feedback frameworks suitable for multiuser systems in  MTC networks.
%
%
First, we  proposed the user {cooperation-based} limited feedback strategy  to obtain high-resolution CSI {with minimal additional burden} on the current FDD-based communication architecture.
We focused on reducing channel quantization errors by allowing {a} limited amount of CSI exchange between close-in users.
In the proposed algorithm, {some}  multiuser resources {are} used {to enhance} channel quantization performance while minimizing  multiuser diversity gain degradation that this approach entails.
Second, we carried out  sum-rate throughput analysis to solve the trade-off problem between  channel quantization performance and   multiuser diversity gain.
Based on the analytical studies, we developed the cooperation mode switching algorithm in order to activate/deactaive  cooperation mode according to channel and network conditions without transmitter interventions.
Numerical results verified that the proposed  algorithm improves   sum-rate throughput because  the multiuser resources  {are used} {to obtain high-resolution CSI.}}


\appendices

\section{Norm of the global effective channel vector}
\label{sec:Appendix_A}

{We study statistical properties of  the \textit{global} channel matrix $\bar{\bH}^{\mathrm{qu}}_a=  [\bH^H_a, \hat{\bh}_{b}^{\mathrm{virt}} ]^H$ in (\ref{eq:virtual_matrix}) to analyze the quantity
\begin{align}
\label{eq:global_effective}
\mathrm{U} \doteq  \| \bar{\bh}_{a|m}^{\mathrm{eff,qu}}\|_2^2=\| (\bar{\bH}^{\mathrm{qu}}_a)^H \bar{\bz}_{a|m}\|_2^2
\end{align}
of the \textit{global} effective channel vector.
The \textit{global} channel matrix is composed of $\bH_a$ including $N$ channel vectors of MU and the (quantized)  virtual channel vector $\hat{\bh}_{b}^{\mathrm{virt}}$  transferred from AU.
It is already known that the entries in $\bH_a$ follow $\mathcal{CN}(0,1)$.}
{We  focus on analyzing  the virtual channel in (\ref{eq:quan_virtual}),
\begin{align*}
\hat{\bh}_{b}^{\mathrm{virt}} &= \cos \phi_{b} \|{\bh}_{b}^{\mathrm{virt}}\|_2    {\bv}_b
\\
&\doteq \cos \phi_{b}  {\bv}_b^{\mathrm{virt}} \in \mathbb{C}^{M},
\end{align*}
where we define ${\bv}_b^{\mathrm{virt}} \doteq \|{\bh}_{b}^{\mathrm{virt}}\|_2    {\bv}_b $. Notice that $\cos\phi_{b}$, $\|{\bh}_{b}^{\mathrm{virt}}\|_2$, and ${\bv}_b$ are independent. 
%
%
Since   $\bv_b$ is an unit-norm vector with isotropically distributed entries,  the  quantity  $\mathrm{A} \doteq  \| {\bv}_b^{\mathrm{virt}}\|_2^2=\|{\bh}_{b}^{\mathrm{virt}}\|_2 ^2$ follows Chi-squared distribution  $\chi^2_{2(M-N+1)}$, where its pdf is given by
\begin{align*}
f_{\mathrm{A}}(a) = \frac{a^{M-N} e^{-a/(2{\sigma^{2}_{a}})} }{{(2\sigma_{a}^2)^{(M-N+1)}}\Gamma{(M-N+1)}}
\end{align*}
with the variance $\sigma_{a}^{2}=\frac{1}{2}$ \cite{Ref_Gor02,Ref_Pro08}.}

{All other entries in the \textit{global} channel matrix  $\bar{\bH}^{\mathrm{qu}}_a$ follow Gaussian distribution so that it would be good  in analyzing the quantity $\| \bar{\bh}_{a|m}^{\mathrm{eff,qu}}\|_2^2$ if the entries in $\bv_b^{\mathrm{virt}}$ also follow  Gaussian distribution.
However, the quantity $\| {\bv}_b^{\mathrm{virt}}\|_2^2$ has the distribution of the $(M-N+1)$-dimensional vector, while the vector is in the $M$-dimensional complex space.
For this reason, ${\bv}_b^{\mathrm{virt}}$ cannot be modeled by Gaussian random variable although its phase follows uniform distribution in the interval $[0, 2\pi)$ and the quantity $\|{\bv}_b^{\mathrm{virt}} \|_2^2$ follows Chi-squared distribution.}
{For easy of analysis, we made two assumptions as follows\footnote{{The accuracy of our approximations are evaluated at the end of Appendix.}}:}
\begin{itemize}
\item {{First, we assume that the quantity $\| {\bv}_b^{\mathrm{virt}}\|_2^2$ can be approximated by $\mathrm{B} \doteq \| \check{\bv}_b^{\mathrm{virt}}\|_2^2$ that follows Chi-squared distribution\footnote{{The accuracy of this approximation increases as the ratio of DoFs between $\mathrm{A}$ and $\mathrm{B}$, i.e.,  $\frac{M-N+1}{M}$, approaches one.}}  $\chi^2_{2M}$ with $M$ degree of freedom (DoF), where its pdf is given by $f_{\mathrm{B}}(b) = \frac{b^{M-1} e^{-b/(2{\sigma^{2}_{b}})} }{{(2\sigma_{b}^2)^{M}}\Gamma{(M)}}$ with the variance $\sigma_{b}^{2}= \frac{M-N+1}{2M} $. The approximation is the result of changing DoF while adjusting its variance to make $\mathrm{E}[\mathrm{B}]$ equal to $\mathrm{E}[\mathrm{A}]$ such that}
\begin{align*}
\mathrm{E}[\mathrm{B}]&= \int_{0}^{\infty} b f_{\mathrm{B}}(b)  \mathrm{d}b=\frac{\int_{0}^{\infty} b^{M} e^{-b/(2{\sigma^{2}_{b}})} \mathrm{d}b}{(2\sigma_{b}^2)^{M}\Gamma{(M)}}
\\
&=\frac{(2\sigma_{b}^2)^{M+1} \Gamma{(M+1)}}{(2\sigma_{b}^2)^{M} \Gamma{(M)}}=M-N+1.
\end{align*}}
%
\item {{Second, we assume that the random variable $\cos^2 \phi_{b}$ can be replaced by its expectation value $\mathrm{E}[\cos^2 \phi_{b}]=1- \omega$ by using similar method\footnote{{It should be pointed out that the variance of the minimum \textit{local} error is very small  compared to that of $\| \check{\bv}_b^{\mathrm{virt}} \|_2^2$. We thus conclude that the operation of replacing the \textit{local} error term by its expectation will lead a minimal effect.}} in Section \ref{sec:cqi_each_cu}. The expectation of the quantization error is derived in Proposition \ref{pr:quantization_error} such that
   $
    \omega ={Q}^{\mathrm{cl}}{\binom{M-1}{N-1}}^{\frac{-1}{M-N}}\mathbf{B}\big({Q}^{\mathrm{cl}},\frac{M-N+1}{M-N}\big).
   $}}
\end{itemize}

{As a result of the above assumptions, the  virtual  vector can be approximated by
$
\hat{\bh}_{b}^{\mathrm{virt}} \simeq \check{\bh}_{b}^{\mathrm{virt}},
$
where its entries follow $\mathcal{CN}\Big(0,\frac{(1- \omega)(M-N+1)}{M}\Big)$.
After all the discussions of the entries in $\bar{\bH}^{\mathrm{qu}}_a$,  the \textit{global} channel matrix  is {modeled} by
\begin{align}
\label{eq:re_re}
\bar{\bH}^{\mathrm{qu}}_a & \simeq \bR^{\frac{1}{2}}\bar{\bH}^{\mathrm{w}}_{a} \doteq \check{\bH}^{\mathrm{qu}}_a
\end{align}
where  $\bar{\bH}^{\mathrm{w}}_{a}$ is the \textit{global} channel matrix having entries {that} follow $\mathcal{CN}{(0,1)}$, and the covariance matrix is given by}
\begin{align}
\label{eq:cova}
\bR  \doteq \left[\begin{array}{cc} \bI_{N} & \mathbf{0}_{N,1}  \\
\mathbf{0}_{1,N} & \frac{(1-\omega)(M-N+1)}{M}\end{array} \right] \in \mathbb{C}^{(N+1) \times (N+1)}.
\end{align}

{The squared norm of the \textit{global} effective channel vector is now approximated by
\begin{align}
\label{eq:global_effective_approx_2}
\| \bar{\bh}_{a|m}^{\mathrm{eff,qu}} \|^{2}_2 & \simeq \frac{\| (\check{\bH}^{\mathrm{qu}}_a)^H \bar{\bu}_{a|m}\|_2^2}{\|\bar{\bu}_{a|m}\|_2^2} = \|\bar{\bu}_{a|m}\|_2^{-2} \doteq \mathrm{D},
\end{align}
because $(\check{\bH}^{\mathrm{qu}}_a)^H\bar{\bu}_{a|m}={\bc}^{\mathrm{proj},m}$, as shown in (\ref{eq:proj_codeword}) and (\ref{eq:combiner_nocoop}).
We derive the distribution of ${\|\bar{\bu}_{a|m}\|_2^{-2}}$.}
\begin{prop}
\label{pr:appendix}
{The distribution of $\|\bar{\bu}_{a|m}\|_2^{-2}$ is the same as the distribution of
$ \frac{\bw^H\bw}{\bw^H\big[\check{\bH}^{\mathrm{qu}}_a (\check{\bH}^{\mathrm{qu}}_a)^H\big]^{-1}\bw}$,
where $\bw \in \mathcal{W}$ is the column vector that is subject to}
\begin{align*}
\mathcal{W} \doteq\Big\{ \bw \in \mathbb{C}^{N+1} :[\bw]_{n}=\frac{e^{j \psi_n}}{\sqrt{N+1}},~\psi_n \sim \mathrm{U}(0,2\pi)\Big\}.
\end{align*}
\end{prop}
\begin{IEEEproof}
{For a given channel matrix $\check{\bH}^{\mathrm{qu}}_a \in \mathbb{C}^{(N+1) \times M}$, the covariance matrix  of $\bar{\bu}_{a|m}$ is computed according to
\begin{align*}
\bZ &\doteq [(\check{\bH}^{\mathrm{qu}}_a)^H]^{\dag} \mathrm{E}[ {\bc}^{\mathrm{proj},m} ({\bc}^{\mathrm{proj},m})^H]  ([(\check{\bH}^{\mathrm{qu}}_a)^H]^{\dag})^H
\\
&\stackrel{(a)}=\frac{\big[\check{\bH}^{\mathrm{qu}}_a(\check{\bH}^{\mathrm{qu}}_a)^H\big]^{-1}\check{\bH}^{\mathrm{qu}}_a(\check{\bH}^{\mathrm{qu}}_a)^H \big[\check{\bH}^{\mathrm{qu}}_a(\check{\bH}^{\mathrm{qu}}_a)^H\big]^{-1}}{M}
\\
&=\frac{ \big[\check{\bH}^{\mathrm{qu}}_a(\check{\bH}^{\mathrm{qu}}_a)^H\big]^{-1}}{M}=\bigg(\frac{ \big[\check{\bH}^{\mathrm{qu}}_a(\check{\bH}^{\mathrm{qu}}_a)^H\big]^{-\frac{1}{2}}}{\sqrt{M}}\bigg)^2,
\end{align*}
where $(a)$ is derived with $\mathrm{E}[ {\bc}^{\mathrm{proj},m} ({\bc}^{\mathrm{proj},m})^H]=\frac{\bI_{M}}{M}$ because ${\bc}^{\mathrm{proj},m}$  is an independent (unit-norm) vector that is isotropically distributed in $\mathbb{C}^M$.
To study statistical distributions of the quantity $\| \bar{\bh}_{a|m}^{\mathrm{eff,qu}} \|^{2}_2$, we model the receive combiner as
\begin{align}
\label{eq:modelled_combiner}
\bu \doteq \bZ^{\frac{1}{2}}\bw,
\end{align}
where $\bw \in \mathcal{W}$ is the isotropically distributed column vector.
By plugging the receive combiner in (\ref{eq:modelled_combiner}) into (\ref{eq:global_effective_approx_2}), the approximated norm square of the \textit{global} effective channel vector is rewritten by
\begin{align*}
\mathrm{D} &\doteq \frac{\| (\check{\bH}^{\mathrm{qu}}_a)^H {\bu}\|_2^2}{\|{\bu}\|_2^2}=\frac{\bw^H \big( \bZ^{\frac{*}{2}} \check{\bH}^{\mathrm{qu}}_a (\check{\bH}^{\mathrm{qu}}_a)^H \bZ^{\frac{1}{2}}\big) \bw}{\bw^H \big(\bZ^{\frac{*}{2}}\bZ^{\frac{1}{2}}\big) \bw}
\\
&=\frac{\bw^H \big( \big[\check{\bH}^{\mathrm{qu}}_a (\check{\bH}^{\mathrm{qu}}_a)^H\big]^{-\frac{1}{2}}\check{\bH}^{\mathrm{qu}}_a (\check{\bH}^{\mathrm{qu}}_a)^H \big[\check{\bH}^{\mathrm{qu}}_a (\check{\bH}^{\mathrm{qu}}_a)^H\big]^{-\frac{1}{2}} \big)\bw}{\bw^H\big[\check{\bH}^{\mathrm{qu}}_a (\check{\bH}^{\mathrm{qu}}_a)^H\big]^{-1}\bw}
\\
&=\frac{\bw^H\bw}{\bw^H\big[\check{\bH}^{\mathrm{qu}}_a (\check{\bH}^{\mathrm{qu}}_a)^H\big]^{-1}\bw}
\end{align*}
and this completes the proof.}
\end{IEEEproof}

{We note that $\check{\bH}^{\mathrm{qu}}_a (\check{\bH}^{\mathrm{qu}}_a)^H \in \mathbb{C}^{(N+1) \times (N+1)}$ is the complex Wishart matrix with the covariance matrix $\bR$ in (\ref{eq:cova}) and $M$ DoF \cite{Ref_Gor02}. Based on  \cite[Theorem 3.2.12]{Ref_Mui09}, we can derive that
\begin{align*}
\mathrm{W} \doteq \frac{\bw^H\bR^{-1}\bw}{\bw^H\big[\check{\bH}^{\mathrm{qu}}_a (\check{\bH}^{\mathrm{qu}}_a)^H\big]^{-1}\bw}= \frac{
\varrho^{-2} \bw^H\bw}{\bw^H\big(\check{\bH}^{\mathrm{qu}}_a (\check{\bH}^{\mathrm{qu}}_a)^H\big)^{-1}\bw}
\end{align*}
follows Chi-squared distribution $\chi^2_{2(M-N)}$, where its pdf is given by $f_{\mathrm{W}}(w) = \frac{w^{M-N-1} e^{-w/(2{\sigma^{2}_{w}})} }{{(2\sigma_{w}^2)^{M-N}} \Gamma{(M-N)}}$ with the variance $\sigma_{w}^{2}=\frac{1}{2}$ \cite{Ref_Gor02,Ref_Pro08}. Note that the numerator  is rewritten by}
\begin{align*}
\varrho^{-2} & \doteq \frac{\mathrm{Tr}(\bR^{-1})}{N+1}=\frac{1}{N+1}\Big[N+\frac{M}{(1-\omega)(M-N+1)}\Big].
\end{align*}

\begin{figure}
\centering
\subfigure{\includegraphics[width=0.435\textwidth]{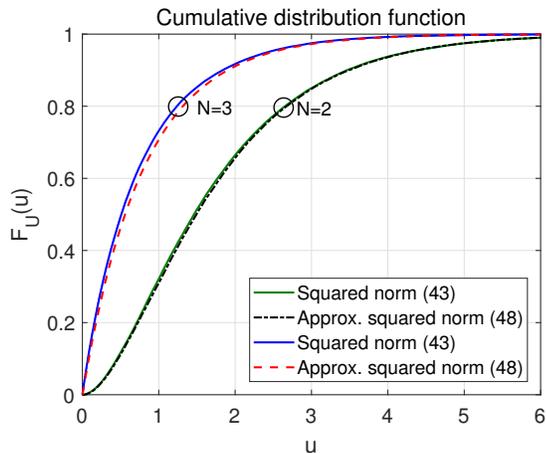}}
\caption{CDF of squared norm of global effective vectors ($M=4$).}
\label{fig:global_vector}
\end{figure}

{The random variable $\mathrm{D}={\|\bar{\bu}_{a|m}\|_2^{-2}}$ in (\ref{eq:global_effective_approx_2}) is  rewritten by
\begin{align}
\label{eq:fin_app}
\mathrm{D}=\varrho^{2}\mathrm{W}=\sum_{\ell=1}^{2(M-N)} \mathrm{D}_{\ell}^{2},
\end{align}
where $\mathrm{D}_{\ell}$ are the i.i.d. entries that follow Gaussian random variables with zero mean and variance  $\frac{\varrho^2}{2}$.
From these observations, the pdf of the quantity $\mathrm{U} \sim \chi^2_{2(M-N)}$  of the \textit{global} effective channel vector can be approximated by}
\begin{align}
\nonumber
f_{\mathrm{U}}(u) &\simeq   \frac{u^{M-N-1} e^{-u/(2(\varrho^2/2))} }{{(2(\varrho^2/2))^{(M-N)}}   \Gamma{(M-N)}}
\\
\label{eq:approximated_pdf}
&= \frac{u^{M-N-1} e^{-u/\varrho^2}}{ \varrho^{2(M-N)} \Gamma{(M-N)}}.
\end{align}

{Finally, we evaluate the approximated \textit{global} effective channel vectors by comparing their squared norm quantities.
Based on the numerical results in Fig. \ref{fig:global_vector}, we conclude that the \textit{global} effective channel vector is well approximated by (\ref{eq:global_effective_approx_2}). For the rest of sections, we use the approximated channel vector with the pdf defined in (\ref{eq:approximated_pdf}).}


\bibliographystyle{IEEEtran}
\bibliography{paper_cooperative}


\end{document}